\documentstyle[12pt]{article}
\begin{document}

\begin{titlepage}
\renewcommand{\thefootnote}{\fnsymbol{footnote}}
\newcommand{\beq}{\begin{equation}}
\newcommand{\eeq}{\end{equation}}
\begin{flushright}
\large
TPI-MINN-94/34-T\\
UMN-TH-1315-94\\
CTP-2391\\
December 1994
\end{flushright}
\vspace*{0.4cm}
\begin{center} \LARGE
{\bf Quantum Mechanics of the Vacuum State in Two-Dimensional
QCD with
Adjoint Fermions}
\end{center}
\vspace*{0.4cm}
\begin{center}

{\bf F. Lenz}

\vspace*{0.1cm}

{\em Institute for Theoretical Physics, University of
Erlangen-N\"{u}rnberg$^{\,\dagger}$,
\\
Staudtstr. 7, 91058 Erlangen, Germany
\\
and
\\
Department of Physics, Massachusetts Institute of Technology,
\\
77 Massachusetts Avenue,
Cambridge, MA 02139-4307, USA}

\vspace*{0.2cm}

{\bf  M. Shifman}

\vspace*{0.1cm}

{\em Theoretical Physics Institute, University of Minnesota$^{\,\dagger}$,
\\
Minneapolis, MN 55455, USA
\\
and
\\
Institute for Theoretical Physics, University of
Erlangen-N\"{u}rnberg
\\
Staudtstr. 7, 91058 Erlangen, Germany}

\vspace*{0.2cm}

{\bf M. Thies}

\vspace*{0.1cm}

{\em Institute for Theoretical Physics, University of
Erlangen-N\"{u}rnberg
\\
Staudtstr. 7, 91058 Erlangen, Germany}

\end{center}

\vskip 4.0cm

$^{\dagger}$ permanent address

\newpage
\vskip 5.0cm

\centerline{{\large \bf Abstract}}
\vskip 1.0cm
A study of two-dimensional QCD on a spatial circle with Majorana
fermions in the adjoint representation of the gauge groups SU(2) and
SU(3) has been performed. The main emphasis is put on the symmetry
properties related to the homotopically non-trivial gauge
transformations and the discrete axial symmetry of this model.
Within a gauge fixed canonical framework, the delicate interplay of
topology on the one hand and Jacobians and boundary conditions arising
in the course of resolving Gauss's law on the other hand is exhibited.
As a result,
a consistent description of the residual $Z_N$ gauge symmetry (for SU(N))
and the ``axial anomaly" emerges. For illustrative purposes, the vacuum of the
model is determined analytically in the limit of a small circle.
There, the Born-Oppenheimer approximation is justified and reduces
the vacuum problem to simple quantum mechanics. The issue
of fermion condensates is addressed and residual discrepancies
with other approaches are pointed out.

\end{titlepage}
\addtocounter{footnote}{0}
\newpage

\section{Introduction}

After the discovery of instantons \cite{Belavin} it became clear that
quantum mechanics of the vacuum state in non-abelian gauge
theories
is non-trivial. Thus, in QCD it was shown
that a non-contractible path in the space of fields exists, i.e.,
one of the directions in the (infinitely dimensional) functional space
is a closed circle. Correspondingly, quantum mechanics of the
vacuum state is analogous to that
of a particle living on a circle. This fact results in a wave function of
the Bloch type and in the occurence of the vacuum angle $\theta$
\cite{Jackiw,Callan}.

At the same time evidence has been mounting that the simple
picture
suggested in Refs. \cite{Jackiw,Callan} is not quite complete, and a
variety
of distinct situations can take place in different theories; in
particular,
the presence of fermions may affect the underlying quantum
mechanics.
Thus, chiral Ward identities in QCD \cite{Crewther} imply that
correlation functions in QCD with massless quarks are  periodic  in
$\theta$ with a period which seemingly depends on the number of
massless
quarks considered
and is not equal to $2\pi$, as one would expect in the
problem of  a particle on  the circle. Another indication of a more
complicated
vacuum structure comes from the four-dimensional supersymmetric
Yang-Mills theories (4-dim SUSYM). Indeed, in such theories
the gaugino condensate $\langle\lambda\lambda\rangle$ is
non-vanishing
and, moreover, exactly calculable \cite{Shifman}. The calculation is
based
on a theorem establishing the holomorphic dependence of
$\langle\lambda\lambda\rangle$ on certain parameters in the
SUSYM
Lagrangian \cite{Shifman2} (for a brief review see the Reprint
Volume
cited in \cite{Belavin}). While the direct calculation \cite{Shifman}
proves that $\langle\lambda\lambda\rangle\neq 0$ the
standard pattern \cite{Jackiw,Callan}
with one circle of unit length in the $K$ direction  ($K$ is the
Chern-Simons
charge; see Fig. 1) gives no hint whatsoever on the
condensation
of $\lambda\lambda$ since the number of the gaugino zero modes
in the tunneling transition with $|\Delta K |= 1$ is larger than two for
any
gauge group.

If for the unitary groups one can at least hope that torons
\cite{tHooft} resolve the paradox  by reducing the length of the
non-contractible contour from 1 to $1/N$ for SU(N)
\cite{tHooft,Witten}, for
the orthogonal and exceptional groups such a way out is absent since
the
torons do not exist in this case. Moreover, the tentative toron solution
of the
problem for the unitary groups does not seem to be appealing since
the
phenomenon of condensation of
$\lambda\lambda$ is quite universal and the condensate
$\langle\lambda\lambda\rangle\neq 0$ is present in the SUSYM
theories
with
arbitrary gauge group.

Thus, it is clear that the existing ideas of quantum mechanics of the
vacuum state in the non-abelian gauge theories are incomplete. The
possibility
of more sophisticated dynamical patterns remains open. Quite
recently it was
shown \cite{Shifman3}, for
instance, that in a ``twisted"  two-dimensional Schwinger model with
two flavors the standard picture of Fig. 1$a$ valid in the one-flavor
model (for a review see e.g.
\cite{Shifman4} ) gives place to that
of Fig. 2. For each given value of $\theta$ there are two vacuum
states.
The two-fold degeneracy and the possibility for the system to live in
the second well, ``half way around the circle", results in the
occurence of a
bilinear fermion condensate which is not generated in the two-flavor
model
otherwise. In a sense, the standard
gauge principle can be extended in the two-flavor model at a price of
incorporating an additional symmetry, which reduces the length
of the non-contractible circle in the space of fields by 1/2.
\footnote{Such an interpretation of the results referring to the
twisted Schwinger model was emphasized by A.Vainshtein.}

In summary, it appears that the standard topological considerations
may not be sufficient to account for all  relevant symmetry
aspects  of gauge  theories; as a consequence the theoretical
analysis has to be performed on a more detailed level for
 achieving a  complete overview of the underlying symmetries.
Such detailed analyses are  possible for a few model
theories only.
Two-dimensional QCD  with fermions in
the adjoint representation of the SU(N) group (QCD$_{2}^{\rm adj}$)
represents such a model. It  is
sufficiently
simple -- transverse degrees of freedom of the gauge fields are
absent, and
we are left with a non-trivial topological structure in its pure form,
not overshadowed by dynamics of ``perturbative gluons".  Moreover,
if the
model is considered on a spatial circle of a small size (as we shall do
in most parts of this paper)
all interesting phenomena take place in the weak coupling regime,
so that quasi-classical methods are fully applicable. This makes the
problem
solvable. On the other hand, the model is sufficiently rich and
some of the dynamical phenomena to be discussed are hopefully
relevant also for  four-dimensional QCD. Most importantly, with
all fields in the adjoint representation, this model exhibits
non-trivial topological properties \cite{Smilga}; the gauge group is
SU(N)/$Z_{N}$
and not SU(N) as is the case for fermions in the fundamental
representation.
Subtle differences in topology of the configurational space can be  the
source
of qualitative
differences in  quantum mechanics of  systems which are similar
otherwise.
This is well-known from other branches of physics; an
example is provided, for instance,  by the so called nematic systems.
The
difference
in topological properties of the magnetic and
nematic
systems leads to distinct quantum mechanics. The configurational
spaces in
these two cases are $S^{2}$ and
$S^{2}/Z_{2}\sim RP^{2}$, respectively; this ``small" distinction
gives rise
to
macroscopically
different properties such as the stability of certain
singularities (line singularities)
in the nematic  but not in the magnetic substances \cite{ROZ}.

Related to the nontrivial topological properties
($\pi_1 [\mbox{SU(N)}/Z_N] = Z_N$) instantons
and bilinear fermion condensates  appear in
QCD$_{2}^{\rm adj}$ \cite{Smilga} much in the same way as in
QCD$_4$.
The fact that QCD$_2$ with adjoint fermions has a kind of $\theta$-vacuum
was first noted in Ref. \cite{Witt79}, where it was also found that the
SU(N) theory has
$N$ vacuum states.
Some other general features of four-dimensional QCD also find
parallels in this two-dimensional gauge theory, as follows from the
recent
analysis  of the large $N$ limit of the model \cite{Klebanov}.

In this work we deal mainly with the gauge groups  SU(2) and
SU(3). It will
be shown that the quantum-mechanical structure of the vacuum
state
in these two cases differs significantly.

We shall investigate QCD$_2^{\rm adj}$ within the
canonical ``gauge fixed" formalism, which is easily derived from
the axial-gauge representation of 3+1 dimensional QCD \cite{LNT}.
Within this formalism, the dynamics is described exclusively in
terms of unconstrained degrees of freedom. As a consequence of the
elimination of redundant gauge degrees of freedom the
topological structure in the space of
gauge fields is implemented in a peculiar although quite explicit
form.
This is generally the case whenever gauge theories are
formulated in
terms of physical variables only. One actually finds in such
unconstrained
formulations residual symmetries, i.e., symmetries which are not
associated
with ``small" gauge transformations. The generators of these residual
symmetries in both abelian and
non-abelian
gauge theories have been completely specified \cite{LNT} -- \cite{LNOT2}
and play a key role in our analysis of quantum mechanics
emerging in QCD$_2^{\rm adj}$.

The questions of gauge fixing and quantization in two-dimensional
QCD were
discussed in the literature more than once \cite{Rajeev} --
\cite{Lang}.
As we will see below, some of the conclusions and results obtained
previously are incorporated in our analysis, while we disagree with
others.

In this work we take up the issue of residual symmetries (as
formulated  in \cite{LNT}) in the specific context of QCD$_2^{\rm
adj}$.
Consideration of this  model helps us  understand, in the language of
quantum mechanics,   implications of the  topological properties.
Furthermore, our analysis  demonstrates a
possibility of revealing symmetries in the process of
elimination of redundant varriables -- symmetries which are seen
in the framework of unconstrained degrees of freedom but are
implicit in the
original formulation.

We work within the  Hamiltonian formalism
which
provides a particularly convenient framework for eliminating the
redundant
gauge degrees of freedom along the lines suggested in Ref. \cite{LNT}
for
QCD$_{4}$; in turn, the results of the present
detailed investigation of QCD$_2^{\rm adj}$ will shed light on general
properties
of such a ``gauge fixed" formalism.
Technically, the
fermions
are assumed to be described by real Majorana fields. Then
QCD$_{2}^{\rm adj}$
is formulated on a finite-size interval (spatial circle), with the
requirement
\begin{equation}
gL\ll 1
\end{equation}
where $g$ is the gauge coupling constant and $L$ is the size of the
interval.
This additional condition ensures that the quantum-mechanical
reduction of
the problem emerging in this way belongs to the weak coupling
regime,
and the structure of the vacuum state can be treated quasi-classically.
The
general strategy is very close in spirit to that accepted previously
\cite{Shifman4}
in the Hamiltonian analysis of the Schwinger model on a circle
(cf. also \cite{MANTON}).
Among other more technical issues  the spectral flow of
the
fermion levels is considered, and the
topological reasons for the  occurence of the fermion zero modes
\cite{Smilga} are explained.

As we will see, the  residual symmetries of the theory can manifest
themselves in fermion
 condensates forming in the corresponding
ground state. The fermion condensates serve as a convenient
indicator of a
vacuum structure.
The question of the fermion condensates in QCD$_{2}^{\rm adj}$ and
a
related
interpretation of the structure of the vacuum state was discussed
recently in
Ref. \cite{Smilga} within the Euclidean (path integral) formulation. If
QCD$_2^{\rm adj}$
is treated on a cylinder $S^1\otimes R$, it has instantons --
trajectories
interpolating
between gauge equivalent points in the space of fields corresponding
to
minima of the effective potential energy. These points are connected
by
gauge
transformations that are not continuously deformable to unity
(although in
this case the
situation is more complicated than just a simple circle of QCD$_4$).
The
instantons generate fermion zero modes. If the gauge group is
SU(2) the
number of zero modes is two -- exactly what is needed to
produce the
bilinear fermion condensate. For higher gauge groups, however, the
number of
the fermion zero modes in the instanton transition is larger than two,
so that
the bilinear condensate does not appear within the standard
instanton
calculus. On the other hand, an independent solution of the model
based on
bosonization seems to show that the bilinear fermion condensate
develops
irrespectively of what particular gauge group is considered. This
paradox, first
mentioned in Ref. \cite{Smilga}, is obviously perfectly identical to the
one
we face in the four-dimensional SUSYM theory.

In our explicit construction of the vacuum wave function the
symmetry properties and related tunneling phenomena are realized
explicitly,
in the familiar context of quantum mechanics of few dynamical
degrees of freedom. Thereby  an intuitive picture of
dynamics  emerges which may be useful in future
attempts to resolve the ``condensate paradox".

The paper is organized as follows. In Sect. 2 we review the
Hamiltonian
approach to QCD$_2^{\rm adj}$. The canonical quantization is carried
out and
the remaining (unconstrained) gauge variables are specified.
The topology of
the corresponding fundamental domains is discussed in detail.
Sect. 3 is devoted to the structure of the vacuum state. The vacuum
wave
function is explicitly built in the limit $gL\ll 1$. The analysis is
conducted
separately for SU(2) and SU(3) theories.
The vacuum wave function is then used for
calculating the fermion condensates.
We also derive
anomaly relations and index theorems relevant to QCD$_2^{\rm adj}$.
Finally we extend our symmetry considerations beyond the weak coupling
approximation. Sect. 4 contains our conclusions and an outlook.

\section{Hamiltonian approach to QCD$_{2}^{\rm adj}$}

\subsection{Formulation of the problem}

The Lagrangian of 1+1 dimensional QCD coupled to Majorana
fermions
$\psi$ in the adjoint representation is
\begin{equation}
{\cal L}= \mbox{tr}\left\{-
\frac{1}{2}F^{\mu\nu}F_{\mu\nu}+i\bar{\psi}D_{\mu}
\gamma^{\mu}\psi\right\}
\label{1}
\end{equation}
with the field strength tensor $F_{\mu\nu}=\partial_\mu A_\nu -
\partial_\nu A_\mu +ig \left[ A_\mu, A_\nu \right]$. The standard
matrix notation is used so that
$$
F_{\mu\nu}\equiv F_{\mu\nu}^a t^a \,\,\, {\rm and} \,\,\, \psi
\equiv \psi^a
t^a
$$
where $t^a$ are the generators of the group, $[t^a,t^b]=if^{abc}t^c$
($t^a =(1/2)\sigma^a$ for SU(2) and
 $t^a=(1/2)\lambda^a$ for SU(3)).
 Moreover, the
covariant derivative acts as
\footnote{If $T^a$ is the generator of the gauge group in the {\em
adjoint}
representation, $(T^a)_{mn} = i f^{man}$, then the
covariant derivative can be written as
$iD_\mu =i\partial_\mu - gA_\mu^aT^a$. Correspondingly,
$F_{\mu\nu}^a =\partial_\mu A_\nu^a -\partial_\nu
A_\mu^a - gf^{abc}A_\mu^b A_\nu^c$. The definitions of the left-
and right-handed fields and $\gamma_5$ below follow
Bjorken and Drell.}
 $$
D_{\mu}=\partial_{\mu} + ig [A_{\mu}, \quad  .
$$
With the following choice of the  gamma matrices
\begin{equation}
\gamma^{0}=\sigma_{2} \,  , \quad \gamma^{1}=i\sigma_{1}\,  ,
\quad
\gamma^{5}= \sigma_{3}
\label{3}
\end{equation}
the Majorana spinors can be chosen real.
In this ``chiral" representation the right-handed fermion field is
the upper component of $\psi$ while the left-handed fermion field is
the lower component.

The theory is considered on a finite-size spatial interval,
$$
0\leq x \leq L \  ,
$$
compactified to a circle by imposing
periodic boundary conditions on the gauge fields,
$$
A_\mu (x=0) = A_\mu (x=L) \, .
$$
As for the fermion fields,
it is slightly more convenient to impose antiperiodic
boundary conditions,
$$
\psi (x=0) = -\psi (x= L)\, .
$$
Before we proceed to quantization let us discuss some general
topological properties
\cite{Smilga,Witt79}
of the theory defined by the Lagrangian (\ref{1}). First of all, since
all fields
considered are in the adjoint representation of the gauge group, the
elements
of the center of  the group act on them trivially. In other words,
the gauge group of the model is not SU(N), but,
rather, SU(N)/$Z_N$. This fact is very important since it leads to a
non-trivial
topology in the space of the gauge fields. Indeed, $\pi_1
[\mbox{SU(N)}]$ is trivial
while $\pi_1 [\mbox{SU(N)}/Z_N] = Z_N$.  The latter aspect reminds
us of
four-dimensional QCD. The parallel is not perfect, though. Indeed in
QCD$_4$ it is
$\pi_3 [\mbox{SU(N)}] = Z $ that counts, and we arrive at one
non-contractible
contour with the topology of a circle (Fig. 1). In other words, if
$U(\vec x )$ is a
gauge
matrix corresponding to the winding number $K=1$ (one full rotation
over the
circle), $U^2$ will represent two rotations, $U^3$ three, etc. In
QCD$_{2}^{\rm
adj}$, say with the SU(2) gauge group,
the analogous matrix $U(x)$ is non-trivial but $U^2$ is already
continuously
deformable to the unit matrix, so that two rotations have the same
effect as
no rotation at all.
Correspondingly, the two-instanton configuration (the $N$ instanton
configuration in SU(N)) is
topologically trivial.

Graphically the topology of the manifold of an analog
quantum-mechanical system might be  depicted as in Fig. 3 (for
SU(2)). The two circles are actually
equivalent to each other, but to make both of them visible they are
split in two and
slightly distorted. The analog particle moves, starting from the point
$O$,
along the larger circle in the direction indicated by the arrow,
approaches $O$
and then moves along the smaller circle in the opposite
direction, ``unwinding" the path. At the point $O$ we, clearly, deal
with a
singularity whose presence seems to invalidate the whole picture.
The general topological arguments do not indicate
 how these properties can  actually be realized in the quantum
mechanical context of the relevant gauge degree of freedom.
Our detailed calculation will unravel the solution to this problem.

  From the general properties of the canonical formalism
the wave function will be shown to vanish at this point
independently of any
details of dynamics. Therefore, the particle motion can be
considered only on one of the two circles of Fig. 3 -- the
configurational space
of the problem is a manifold {\em with boundaries}. This manifold
will be
referred to as the fundamental domain and will be discussed in more
detail in
Sect. 2.3. Here we only note that dynamics on the second circle
of Fig. 3 is not independent and is just a gauge replica of that of the
fundamental domain.

If a single Majorana field in the adjoint representation is considered
-- as we shall do here --
such a theory has no continuous fermion symmetries whatsoever.
Indeed the real fermion field does not allow any phase rotation,
neither
for the left-handed nor for the right-handed components. There exist
no
conserved gauge invariant (colorless) fermion currents since
\begin{equation}
\mbox{tr} (\bar\psi\gamma_\mu\psi )\equiv 0
\,\,\, {\rm and}\,\,\,
\mbox{tr} (\bar\psi\gamma_\mu\gamma_5\psi )\equiv 0 \ .
\label{currents}
\end{equation}
We shall see later, however, that
one can define a colored vector fermion current which, in a {\em
certain
gauge}, is conserved in the normal sense -- its regular, not covariant
divergence, vanishes. The divergence of its
$\gamma_\mu\gamma_5$ partner
has an
anomaly looking similar to the axial anomaly in the Schwinger model
(in the limit $gL \ll 1$).

With $n$ Majorana fields the Lagrangian of the model
would possess a global O(n)$\times$O(n) symmetry. Any bilinear
fermion
condensate would necessarily break a part of this symmetry
spontaneously.
Since in two dimensions spontaneous breaking of continuous
symmetries cannot
take place, such a model would be protected from the generation of
bilinear condensates and, although essential features of the dynamics
remain
intact, the analysis would become more involved. This is the reason
why
for the time being we focus on the one-flavor case. Our model is
engineered to
be a prototype of QCD$_4$ with one massless quark. Multi-flavor
generalizations would be closer to QCD$_4$ with two or more
massless quarks.

\subsection{Canonical quantization}

We now proceed to the quantization of the system described above.
Since as in any gauge theory we deal here with a large number of
redundant
(gauge) degrees of freedom  two alternative aproaches might be
applied. One possibility is eliminating the gauge degrees of freedom
at the
classical
level and then quantizing the remaining physical degrees of freedom.
This is the most commonly used procedure which however seems
to be fraught with certain ambiguities   concerning the kinetic energy
of
the gauge fixed degrees of freedom (cf. Refs. \cite{Hetrick},
\cite{LTLY} versus Ref. \cite{LNT}).
 The other possibility is to quantize the system in the Weyl gauge
\begin{equation}
A_{0}=0
\label{4}
\end{equation}
with part of the redundant variables
(the space components of $A_{\mu}$) subject to constraints, and to
resolve the constraints at the quantum level.
This second alternative has the attractive feature that the quantization is
completely straightforward.
We shall explain in detail how this works in the present case
and how we deal with the constraints. Our final results will shed light
on the consequences associated with the ambiguities of the
``classical"
gauge fixing procedure and thereby help to resolve these issues.

In the Weyl gauge, the Gauss law is not an equation of motion but
has to be
imposed ``by hand". Canonical quantization is straightforward,
yielding $N^{2}-1$ gauge field variables $A_{1}^{a}$ and their
conjugate variables, the
electric fields $E^{a}$.
The Hamiltonian for QCD$_{2}^{\rm adj} $ is
\begin{equation}
H=\int_{0}^{L} dx \,\,  \mbox{tr}\left\{\frac{1}{2}E^{2}-
i\psi\sigma_{3}D_{1}\psi\right\} \ .
\label{5}
\end{equation}
The Gauss law is implemented as a constraint on the
physical states,
\begin{equation}
\left( \partial_{1} E(x)-ig \left[A^{1}(x),E(x) \right] - g\rho(x) \right)
|\Phi \rangle =
\left( D_{1} E(x)
-g\rho (x)\right) |\Phi\rangle =0 \ .
\label{6}
\end{equation}
Here $\rho$ is the fermion  color charge density  in the matrix
representation,
\begin{equation}
\rho^{ij}\equiv \rho^a (t^a)_{ij} = \frac{1}{2}\left(
\psi^{ik}\psi^{kj} -\psi^{kj}\psi^{ik}
\right)
\label{7}
\end{equation}
where the summation over the Lorentz indices of $\psi$ is implied.

In this framework, the technically demanding part is not the
quantization but the resolution
of Gauss's law. In 2-dimensional QCD, this issue is particularly crucial
since
the dynamics of the gauge field is almost exclusively governed by
the Gauss law constraint.
Various techniques to resolve Gauss's law have been advocated in
the literature.
Recently, a new method based on unitary transformations in the
Hilbert
space (rather than on a change of variables in the Schr\"{o}dinger
representation \cite{Christ}) has been developed \cite{LNOT1,LNOT2}
and successfully applied to 4-dimensional QCD on a torus, in a
modified axial gauge \cite{LNT}.
Since the axial gauge singles out one space direction,
the results of Ref. \cite{LNT} can easily be adapted for  QCD$_2$.
Without repeating all  details, we briefly sketch
this ``gauge-fixing" procedure and collect the pertinent results.

In the first step, one tries to resolve the Gauss law constraint
(\ref{6}) for $E(x)$
and to substitute the corresponding expression for $E(x)$ in the
Hamiltonian (\ref{5}), restricting
oneself from the outset to the physical sector. This requires inversion
of the covariant derivative $D_{1}$
viewed as operator in color and configuration space,
which is most conveniently done after diagonalization.
One finds that $D_{1}$
has generically (i.e., for arbitrary $A^{1}$) $N-1$ zero modes. The
projections of $E$ onto these zero modes, denoted by
$\frac{1}{L}e^{p}, p=1,\ldots ,N-1$,
are obviously not constrained by Gauss's law and remain as physical
variables in the
Hamiltonian; however, due to the use of a dynamical basis \ -- the
eigenfunctions of $D_{1}$ depend on the gauge field $A^{1}$ -- \  they
are non-hermitian
quantum-mechanical operators.
The projections of $E$ onto the non-zero modes of $D_{1}$ on the
other hand can be eliminated in the space of physical states
in favour of the (matter) color charge density, $\rho$,
and $A^1$ (through the inverse of the operator $D_{1}^{\prime}$,
where the prime means that zero modes are omitted); schematically:
\begin{equation}
\langle \Phi | E^{a}E^{a} |\Phi \rangle = \langle \Phi | \left(
\frac{1}{L^{2}} e^{p \dagger} e^{p}
+  2 \mbox{tr}\,  g\rho \frac{1}{D_{1}'} g\rho  \right)    | \Phi
\rangle\, .
\label{7a}
\end{equation}
Note that here and throughout this paper, we shall use the letters
$a$, $b$,
etc. belonging to the beginning of the alphabet for the $N^{2}-1$
generators
of the algebra,
while those of the Cartan subalgebra will be denoted by
$p$, $q,\ldots .$ The corresponding summation conventions are
implied in equations such as eq. (\ref{7a}).

At this stage, the Hamiltonian still contains
the full gauge field $A^1$.
However, one can construct a unitary transformation which
eliminates
all modes of $A^{1}$ except for $N-1$ spatially constant quantum
mechanical variables $a^{p}$, related to the eigenvalues of the path
ordered exponential winding around the
circle,
\begin{equation}
 \mbox{P} e^{ ig\int_0^L \, dx A^1(x)}
= V e^{igaL}V^{\dagger} \ , \ \ \ \ \left(a= a^{p}t^{p}\right) \ .
\label{7b}
\end{equation}
This procedure is satisfactory beacuse
one can show that the constant electric fields $e^{p}$ (projection of
$E$
onto the zero modes of $D_{1}$) and the zero mode ``gluons" $a^{q}$
(closely related to the eigenvalues of $D_{1}$)
are conjugate variables satisfying
standard commutation relations
\begin{equation}
[e^{p},a^{q}]=\frac{1}{i}\delta_{pq} \quad
p,q=1,\ldots,N-1 \ .
\label{9}
\end{equation}
The resulting theory still has global constraints: The projection of
the Gauss law onto the zero modes of $D_{1}$ yields $N-1$
residual, $x$-independent constraints which  --  after the gauge
fixing unitary transformation  --
are reduced to the condition that the neutral fermion charges vanish,
\begin{equation}
Q^{p} |\Phi \rangle = \int_{0}^{L} dx \rho^{p}(x) |\Phi \rangle = 0
\quad \quad
(p=1, \ldots ,N-1) \ .
\label{9b}
\end{equation}
One characteristic feature of this method is the fact that $e^{p}$ is not
hermitian, although the
Hamiltonian is. This is completely analogous to the transition from
cartesian to polar coordinates in quantum mechanics where
$e^{p}$ would
correspond to the non-hermitian radial momentum operator
$\hat{\vec{r}}\cdot \vec{\nabla}/i$. In both cases, a hermiticity
defect arises
when projecting the momentum operator on a dynamical, i.e.,
coordinate
dependent ($A^1$ and $\vec{r}$, respectively), basis.
As in the  familiar case of the polar coordinates,
the kinetic energy can be rewritten with
the help of an appropriate Jacobian in the form
\begin{equation}
K_{a} = \frac{1}{2L}  e^{p \dagger} e^{p} =
-\frac{1}{2L}
\frac{1}{J\left[a\right]}\frac{\partial}{\partial
a^{p}}J\left[a\right]\frac{\partial}{\partial a^{p}} \ .
\label{10a}
\end{equation}
The Jacobian
$J[a]$, in turn, can be readily deduced from the non-hermitian part
of
$e^{p}$. In
the SU(N) case one finds
that $J[a]$ is just the Haar measure of this group,
\begin{equation}
J[a]=\prod_{i>j}\sin^{2}\left(\frac{1}{2}gL(a_{i}-a_{j})\right)\, ,
\label{10b}
\end{equation}
$$
a_{i} = a^{p}t^{p}_{ii} \  \quad (\mbox{no summation over}\,\,  i)\ ,\quad
\sum_i
a_{i} = 0  \,  ,
$$
reflecting the transition from elements of the algebra (vector
potentials) to elements of the group (path ordered exponentials)
in the gauge fixing process.

Eventually, we arrive
at the following Hamiltonian describing the dynamics of the gauge
variables
$a^p$ and the fermion degrees of freedom in the space of physical
states,
\begin{equation}
H=K_{a} +H_{C} +H_{\rm ferm} \ .
\label{13}
\end{equation}
$K_{a}$ is the kinetic energy of the gauge degrees of freedom, eq.
(\ref{10a}).
$H_C$ is
the color electrostatic energy (``Coulomb energy") appearing in the
elimination of the
electric field by means of the Gauss law, cf. the 2nd term on the r.h.s.
of eq. (\ref{7a}),
\begin{equation}
H_{C}=\frac{g^{2}}{L}\sum_{n=-\infty}^{\infty}\sum_{i,j}\int_{0}^{L}
dy\int_{0}^{L}
dz\left(1-\delta_{ij}\delta_{n 0}\right)
\frac{\rho^{ij}\left(y\right)\rho^{ji}\left(z\right)}{\left(\frac{2\pi
n}{L}+
g\left(a_{i}-a_{j}\right)\right)^{2}}e^{2i\pi n\left(y-z\right)/L} .
\label{16}
\end{equation}
The fermionic part of the Hamiltonian has the form
\begin{equation}
H_{\rm ferm}=-i\int_{0}^{L} dx \,\,  \mbox{tr}\,
\left\{\psi\sigma_{3}\left(\partial_{1}\psi-ig [a,\psi] \right)
\right\} .
\end{equation}
The fermion degrees of freedom are quantized in a standard way via
\begin{equation}
\left\{ \psi_\alpha^{ij}(x), \psi_\beta^{kl}(y)\right\}
=\frac{1}{2}
\delta_{\alpha\beta}\delta^{(a)} (x-y)\left(
\delta_{il}\delta_{kj} - (1/N)\delta_{ij}\delta_{kl}\right)
\label{antic}
\end{equation}
where the $\delta$-function $\delta^{(a)}(x-y)$ is the one
appropriate for an interval of length $L$ and anti-periodic boundary
conditions.

It is convenient to classify all fermion fields with respect to color
degrees of freedom in the following way. First, we single out the
``neutral" components of the fermion field, $\psi^p$, $p=1,..., N-1$.
It is obvious that they
are not coupled to the gauge degrees of freedom
$a^p$. They only take part in the Coulomb interaction which is
suppressed in
the small interval limit.

The off-diagonal components of the fermions
are charged with respect to the neutral zero-mode ``gluons" $a^p$.
We
introduce
$\frac{1}{2}N(N-1)$ charged fields
\begin{equation}
\varphi^{ij}=\sqrt{2} \psi^{ij} \,  , \quad \varphi^{ij
\dagger}=\sqrt{2} \psi^{ji}
\,   \quad\mbox{(for}\, \,\,  i<j \mbox{)}\, ,
\label{11}
\end{equation}
satisfying  the standard canonical anticommutation relations
(cf. eq. (\ref{antic}))
\begin{equation}
\left\{\varphi^{ij}_{\alpha}\left(x\right),\varphi^{kl
\dagger}_{\beta}\left(y\right)\right\}=\delta_{\alpha\beta}\delta_{i
k}\delta_{jl}\delta^{(a)}\left(x-y\right) \ .
\label{12}
\end{equation}
With these definitions
the fermion part of the  Hamiltonian takes the form
\begin{equation}
H_{\rm ferm} = H_\psi + H_\varphi(a) ,
\end{equation}
where
\begin{equation}
H_{\psi}= \frac{1}{2i} \int_{0}^{L} dx
\psi^{p}\sigma_{3}\partial_{1}\psi^{p} ,
\label{14}
\end{equation}
and
\begin{equation}
H_{\varphi}\left(a\right)=\sum_{i<j}\int_{0}^{L} dx \varphi^{ij
\dagger}\sigma_{3}\left(\frac{1}{i}\partial_{1}-g(a_{i}-
a_{j})\right)\varphi^{ij} .
\label{15}
\end{equation}
The theory is supplemented by the neutrality conditions (\ref{9b})
which
in our present notation read
\begin{equation}
Q_{i}|\Phi\rangle = 0 \ , \ \ \quad i=1,\ldots ,N-1
\label{20}
\end{equation}
where (cf. eqs. (\ref{7}), (\ref{11}))
\begin{equation}
Q_{i}=\int_{0}^{L} dx \rho^{ii}=\frac{1}{2}\int_{0}^{L} dx
\left(\sum_{k>i}\varphi^{ik \dagger}\varphi^{ik}-
\sum_{k<i}\varphi^{ki
\dagger}\varphi^{ki}
\right)
\label{21}
\end{equation}
(no summation over $i$).

The appearance of the Jacobian
(cf. eqs. (\ref{10a}), (\ref{10b}))
in the kinetic energy of the quantum-mechanical gauge degrees of
freedom,
will turn out to be vital for the self-consistency of our
quantum-mechanical solution of the problem.
The eigenvalue problem with the
Hamiltonian $H$ defined in eqs. (\ref{10a}), (\ref{13}), (\ref{16}),
(\ref{14}) and (\ref{15})
has the form
$$
H|\Phi\rangle  = E|\Phi\rangle
$$
where $|\Phi \rangle $ denotes a state vector in the physical space.
We can perform a similarity transformation and proceed
to a different wave function which, for brevity, will be referred to as
``radial'',
\begin{equation}
\tilde{\Phi}[a]=\sqrt{J[a]}\Phi [a] .
\label{18}
\end{equation}
In the space of the ``radial" wave functions,
the kinetic energy operator acts as follows,
\begin{equation}
- \frac{1}{2L} \frac{1}{\sqrt{J}} \partial_{p}
J \partial_{p} \frac{1}{\sqrt{J}} = - \frac{1}{2L}
\partial_p \partial_p +
\frac{1}{2L} \frac{1}{\sqrt{J}}\left(\partial_p \partial_p \sqrt{J}
\right)
\label{15a}
\end{equation}
(with the notation $\partial_{p} = \partial/\partial a^{p}$),
i.e., as the standard Laplace operator supplemented by an ``effective
potential". In the present case,
this effective potential turns out to be a constant
\begin{equation}
\frac{1}{2L} \frac{1}{\sqrt{J}}\left(\partial_p \partial_p \sqrt{J}
\right)
= - \frac{(gL)^{2}}{48} N (N^{2}-1)
\label{15b}
\end{equation}
which can simply be dropped since it merely shifts all energy
eigenvalues by the same amount.
The only reminiscence of the Jacobian will then be the boundary
condition
for the radial wave function
\begin{equation}
\tilde{\Phi}[a]=0 \, \ \ \mbox{if} \quad J[a]=0 \ ,
\label{19}
\end{equation}
analogous to the condition that the radial wave function vanishes at
$r=0$
when working with polar coordinates. We emphasize that in
contradistinction to
the case of the polar coordinates, the Jacobian defines compact
intervals (domains)
and, therefore, guarantees the existence of discrete spectra and
normalizable
wave functions.

We also note the correspondence between the electrostatic field
energy $H_{C}$ and
the
centrifugal barrier in the conventional radial Schr\"{o}dinger
equation.
Both are singular at the points where  the respective
Jacobians have zeroes. Indeed, the Coulomb interaction part $H_C$,
eq. (\ref{16}), is
formally
ill-defined whenever $a_{i}-a_{j} = 2\pi n/gL$, exactly where the
Jacobian
(\ref{10b}) vanishes.

Before discussing the symmetries of this theory, let us briefly
comment on the
alternative quantization scheme mentioned at the beginning of this
section (``complete" classical gauge fixing followed by quantization).
Here one starts from the observation
that all non-constant modes of $A^1$
can be ``gauged away". The remaining constant modes
of $A^1$, by  global rotations in the color space, can be reduced to a
diagonal matrix.  Thus, the physical degrees of freedom in
$A^1$ are the constant modes of $A^{1,p}$ (where the index $p$ runs
again over $N-
1$ values corresponding to
the Cartan subalgebra of SU(N)).
Moreover, $A_0$, the time component of the gauge potential, enters
the
Lagrangian without time derivatives. This means that it is not
dynamical
and can be eliminated altogether by means of the equations of
motion.
In this process, the Coulomb interaction is generated.

This scheme has been followed, for example, in Ref. \cite{LTLY} (in
light-cone coordinates, but
this is immaterial for the present discussion of gauge fixing).
For the Schwinger model,
the resulting formulation is indistinguishable from the one derived
via the Weyl gauge.
For QCD$_{2}$ however, the Jacobian was missed, so that the
Hamiltonian looks
exactly like the one above (after going to the radial wave function
and the standard form of the kinetic energy),
but without the boundary condition (\ref{19}) and the effective
potential.
We conclude that this framework is not yet completely defined,
unless one is
able to recover the Jacobian from some other source.

The origin of this discrepancy can easily be singled out: In QED$_{2}$,
the physical (gauge invariant) variable on the circle is the zero mode
of $A^{1}(x)$,
\begin{equation}
a=\frac{1}{L} \int_{0}^{L} dx A^{1}(x) \ .
\label{19b}
\end{equation}
Therefore, it is sufficient to simply discard the non-zero modes of $A^1$
as one would do in naively fixing the gauge to
$\partial A^{1}/\partial x
= 0$.
In QCD$_{2}$,
the correct gauge invariant variables are not the standard zero
modes of $A^{1}$,
but rather the elements of the diagonal matrix $a$ defined as (cf. eq.
(\ref{7b}))
\begin{equation}
a = \frac{1}{igL} \log \left( V^{\dagger} \mbox{P}  e^{ig\int_{0}^{L} dx
A^{1}(x)} V \right) \ .
\label{19c}
\end{equation}
One cannot keep track of the difference between eqs. (\ref{19b}) and
(\ref{19c})
--  which is responsible for the non-trivial Jacobian in the QCD case --
if one simply
drops certain color and Fourier components of the $A^1$-field at the
classical level.
In order to do it correctly, one presumably would have to implement
the gauge
condition classically as a change of variables, in a Hamiltonian
framework, and then quantize the theory in curvilinear coordinates.
The fact that at least in the present example one only
misses a boundary condition suggests that there might be some
short-cut,
but we are not aware of it. In any case,
the conceptual advantage of the Weyl gauge is so
tangible that it seems the most natural method within the canonical
framework.
Moreover, the technique of resolving Gauss's law via unitary
transformations
has already proven to be helpful in clarifying the physical meaning
of the
residual gauge symmetries in the case of abelian theories
\cite{LNOT1,LNOT2}.

It is in order to note that the Hamiltonian emerging in this way in SU(2)
QCD$_2^{\rm adj}$ after the transformation
(25) is {\em locally} indistinguishable from the Hamiltonian one gets
in the
Schwinger model after  the gauge fixing procedure (provided that we
neglect,
for the time being, the term $H_C$ which is inessential in the
limit $gL \ll 1$).
The ``analog" Schwinger model we end up
with
is the model of one  Dirac fermion  with unit charge coupled to a
neutral
``photon" $a$. Similarly, the SU(3) case looks locally like a generalization
of the Schwinger model to the gauge group U(1)$\times$U(1); here, two neutral
``photons",
$a^3$ and $a^8$, are coupled to three Dirac fermions,
$$
\varphi^{12} ,\quad \varphi^{13} \quad\mbox{and}\,\,\,
{\varphi}^{23} .
$$
The $a^3$ charges of these fermions are $1$, $1/2$ and $-1/2$,
respectively,
while the $a^8$ charges are $0$, $\sqrt{3}/2$ and $\sqrt{3}/2$. The
distinction between
QCD$_2^{\rm adj}$ and the Schwinger model
-- which is absolutely crucial -- shows up only in the global
properties:
namely, the domain where the variables $a$ live, how different
points on
the corresponding manifolds are glued, etc. We proceed to a discussion
of this
issue.

\subsection{Fundamental domain, residual symmetries and large
gauge transformations}

In the previous section the dynamics of the gauge fields was reduced
to that of the residual gauge variables $a^p$ by resolving Gauss's law.
Let us first discuss the range of values these variables can take on.
In the process of gauge fixing,
the $a^{p}$ are only defined via group elements, namely the
eigenvalues
of the path ordered integral around the circle, cf. eq. (\ref{7b}).
Therefore, $gLa_{i}$ are angular variables, which are only defined
modulo $2\pi$.
If we impose two conditions:
\begin{itemize}
\item the parametrization is one-to-one,
\item permutations of the eigenvalues do not lead out of the domain,
\end{itemize}
the definition of such a domain for the variables $a^{p}$ is unique.
We will call
this domain {\em the elementary cell}.
For SU(2), it is the interval
\begin{equation}
-\frac{2\pi}{gL} \leq a^{3} \leq \frac{2\pi}{gL}
\label{19e}
\end{equation}
with the endpoints identified (Fig. 4a). The points where two eigenvalues of
the matrix eq. (\ref{7b}) cross are
0 and $\pm \frac{2\pi}{gL}$, therefore exchanging $a_{1}$ with
$a_{2}$ maps the two
half intervals onto each other, $a^{3} \to -a^{3}$ (recall that $a_1 =a^3/2,
\,\,\,  a_2 =
-a^3 /2 $).

For SU(3), the corresponding construction can easily be found with
the help
of the standard Gell-Mann matrices $\lambda^{3}$ and
$\lambda^{8}$ which yield
\begin{equation}
a_{1}  =  \frac{1}{2} \left( a^{3} + \frac{1}{\sqrt{3}} a^{8} \right)\ ,
\quad
a_{2}  =   \frac{1}{2} \left( -a^{3} + \frac{1}{\sqrt{3}} a^{8} \right)\ ,
\quad
a_{3}  =   -\frac{1}{\sqrt{3}} a^{8}  \ .
\label{19f}
\end{equation}
Here, 2 out of the 3 eigenvalues of the matrix eq. (\ref{7b}) cross
along the lines
\begin{equation}
\frac{gLa^{3}}{2} = n \pi \ , \quad \pm \frac{gLa^{3}}{2}+ \sqrt{3}
\frac{gLa^{8}}{2} = 2n\pi   \ , \quad(n \in Z)
\label{19g}
\end{equation}
At the points of intersection of these lines, all 3 eigenvalues are
equal.
The resulting elementary cell is the hexagon shown in Fig. 4b in the
$\{ a^{3},
a^{8}\} $ plane.
If we tile the plane with such hexagons, then shifting any of the 3
angles $gLa_{i}$ by $2\pi$
maps this domain onto another such hexagon. Our choice is singled
out by the condition that permutations of the eigenvalues just map
the different
sub-triangles onto each other, so that the full domain is invariant.
Note
that opposing sides of the hexagon should be identified since they
correspond to angles differing by $2\pi$.

Construction of the elementary cell is not the end of the story,
however.
We could have factorized it with respect to the Weyl group. In other
words,
we still have the freedom of performing (topologically trivial) gauge
transformations which will order the eigenvalues. For instance, in the
SU(2)
case, if $a^3$ is negative, we can perform a global rotation in the
color space,
by $\pi$ around the first axis. This rotation maps the interval
$-2\pi/(gL)<a^3<0$ onto the interval $0<a^3<2\pi/(gL)$
(simultaneously,
$\varphi^{12}\leftrightarrow\varphi^{21}$).
In this way we arrive at the notion of
fundamental domain --
a domain of variation of $a$'s emerging after we use all gauge
freedom residing in the topologically trivial gauge
transformations. Further reduction of this domain
(identification of points that are gauge images of one another)
is possible only by invoking ``large" (topologically non-trivial)
gauge transformations.

Generically,
if  we specify the gauge  by
prescribing
the ordering of the eigenvalues, we  restrict ourselves to the
fundamental domain smaller than the elementary cell by a factor of
$(1/N!)$,
namely half
the interval in SU(2) and one equilateral triangle in SU(3).
In fact, these fundamental domains are separated by points (in
SU(2))
or lines (in SU(3)) of vanishing Jacobian, hence they are anyway
dynamically decoupled.
This singles out these smaller, {\em fundamental} domains as the
relevant ones, and
we choose them as (see Fig. 4)
\begin{eqnarray}
0 \leq & a^{3} & \leq \frac{2\pi}{gL} \quad \quad \quad \ \ \,
\mbox{in} \ \mbox{SU(2)} \ \mbox{and} \ \mbox{SU(3)} \nonumber
\\
-\frac{1}{\sqrt{3}} a^{3} \leq & a^{8} & \leq +\frac{1}{\sqrt{3}}a^{3}
\quad \quad \mbox{in} \ \mbox{SU(3)}
\label{19ga}
\end{eqnarray}

We now turn to the formal symmetries of our Hamiltonian. For this
discussion
it is actually advantageous  not to restrict the range of variables as
described above but rather work with variables of the covering
space.
As shown in Ref. \cite{LNT},
these symmetries  are the left-overs of the original local gauge
symmetry
in a gauge-fixed
formulation. Since these residual symmetries
can easily be verified in the present case,
we only summarize the results:
\begin{itemize}
\item Displacements  $T_{D}(\vec{k})$
\begin{eqnarray}
a_{i} & \to & a_{i} + \frac{2\pi}{gL} k_{i} \quad \quad \left(\sum_{i}
k_{i}  = 0, k_{i} \in Z\right) \nonumber \\
\varphi^{ij} & \to & e^{2\pi i (k_{i}-k_{j})x/L} \varphi^{ij}  \ , \quad
\psi^{p}  \to \psi^{p}
\label{19h}
\end{eqnarray}
\item Central conjugations $T_{C}(n)$
\begin{eqnarray}
a_{i} & \to & a_{i} + \frac{2\pi}{gL} \nu_{i} \quad \quad (\nu_{i} =
n(1/N -\delta_{iN}),\  n= 1 , \ldots , N-1) \nonumber \\
\varphi^{ij} & \to & e^{2\pi i (\nu_{i}-\nu_{j})x/L} \varphi^{ij} \ ,
\quad
\psi^{p}  \to \psi^{p}
\label{19i}
\end{eqnarray}
\item Permutations of the color basis $(P(1)...P(N))$
\begin{equation}
a_{i}  \to  a_{P(i)}  \ ,  \quad
\psi^{ij}  \to  \psi^{P(i)P(j)}
\label{19j}
\end{equation}
\item Global, color diagonal transformations
\begin{equation}
a_{i}  \to  a_{i} \ , \quad
\varphi^{ij}  \to  e^{i(\beta_{i}-\beta_{j})} \varphi^{ij} \ , \quad
\psi^{p} \to
\psi^{p}
\label{19k}
\end{equation}
\end{itemize}
where $\beta_i$ are real numbers.
The permutations (\ref{19j}) are awkward to write down in the basis
of
$\psi^{p}, \varphi^{ij}$
for general $N$, so that we have used here the original fermion fields
$\psi^{ij}$ instead.
For SU(2) and SU(3), the conversion can easily be made if needed.
The global symmetry (\ref{19k}) reflects the fact that the residual
neutrality
condition (\ref{20}), the remnant of the original Gauss law constraint,
 has not yet been implemented, and trivially reduces to {\bf 1}
in the space of physical states.

It is instructive to confront these formal symmetries of the system
when using
variables in the covering space with the choice
of the fundamental domain. The displacements (\ref{19h}) shift the
fundamental domain to some other possible domain, since an angle
$gLa_{i}$ gets shifted
by $2\pi$. The permutations have already been discussed. The
central
conjugations amount to a shift of $a^{3}$ by $2\pi/(gL)$ (half the
length of the elementary cell)
for SU(2), and to a shift of $a^{8}$ by $4\pi /(\sqrt{3}gL)$
(half the ``diameter" of the hexagon) for SU(3). Clearly,
restriction of the variables to the fundamental domain
renders most of these symmetries not
very useful at this stage. How this is
resolved can best be explained with the help of the concrete
examples below.

As last item, we discuss the large gauge transformations and their
relation
to the symmetries of our Hamiltonian.

First consider SU(2).
As discussed in subsection 2.1, the elements of the
center of the group act trivially if the fermions are
in the adjoint representation of the group. The gauge group
is actually not SU(2) but, rather, SU(2)/$Z_2$ (or SO(3)). Thus, we
need to discuss
mappings of the (spatial) circle onto SU(2)/$Z_2$. As is well-known,
$\pi_1 [\mbox{SU(2)}/Z_2] = Z_2$, i.e., there exist matrices $U(x)$
defined on the
circle and not continuously deformable to the unit matrix. It is easy
to
construct an example of such a matrix. Indeed, the group space of
SU(2)/$Z_2$ is
the 3-sphere, $S^3$, with the endpoints of any diameter identified. A
mapping of the
circle starting from the south pole of the sphere, going along a
meridian and
ending up at the north pole is, therefore, allowed and it obviously
cannot be continuously deformed to the trivial mapping, for example
\begin{equation}
U(x) = {\rm e}^{i\frac{\pi x}{L}\sigma^3} \ .
\label{su2u}
\end{equation}
More generally, all $U(x)$ can be classified according to the relation
\begin{equation}
U_{\pm}(L)=\pm U_{\pm}(0) \ .
\label{19ka}
\end{equation}
All transformations $U_{-}$ are topologically non-trivial, but can be
deformed into
each other by ``small" gauge transformations $U_{+}$.
The square of any $U_{-}$ clearly belongs to the topologically trivial
class.

It is obvious how to generalize this to SU(N): The gauge group is
SU(N)/$Z_N$.
Topologically non-trivial are just the gauge transformations differing
by a non-trivial center element
if one goes around the circle,
\begin{equation}
U_{z}(L) = z U_{z}(0)   \ ,
\label{19l}
\end{equation}
where $z$ is the $n$-th root of unity ($\neq 1$),
\begin{equation}
z=-1 \ \ \mbox{for} \ \ \mbox{SU(2)}\ ,\ \  z={\rm e}^{i2\pi/3}\ ,\
{\rm e}^{i4\pi/3}\ \  \mbox{for} \ \ \mbox{SU(3)}  \ .
\end{equation}

How does $a^{3}$ behave under these gauge transformations?
If we go back to the defining relation (\ref{7b}) and perform a local
gauge transformation $U(x)$, we get
\begin{equation}
 \mbox{P} e^{ig\int_{0}^{L}dx A^{1}(x)} \to
U(L) \mbox{P} e^{ig\int_{0}^{L}dx A^{1}(x)} U(0) = U(L) V e^{igaL}
V^{\dagger} U(0) \ .
\label{19kb}
\end{equation}
Together with eq. (\ref{19ka}), this yields
\begin{equation}
V e^{igaL} V^{\dagger} \to \pm U_{\pm}(0) V e^{igaL} V^{\dagger}
U^{\dagger}_{\pm}(0) \ ,
\label{19kc}
\end{equation}
where the $+$ sign holds for small, the $-$ sign for
large gauge transformations. It is tempting to conclude from this that
$a^{3}$ gets shifted by
$\frac{4\pi}{gL}$ resp. $\frac{2\pi}{gL}$. The detailed
transformation property depends on the precise definition of the
fundamental
domain.
Since one only knows how the exponential transforms, one cannot
infer from
that how the exponent transforms, unless one fully specifies
how to take the logarithm. Furthermore, the ordering of the eigenvalues
is involved.
Fortunately, it will turn out that the present formalism projects out
the relevant symmetry without the necessity of submerging into
these subtle issues.  Moreover, the gauge transformations glueing the
points at the boundaries of the fundamental domains are easily
identifiable both, for SU(2) and SU(3).

Repeating a similar analysis for SU(N) yields the transformation
property
\begin{equation}
V e^{igaL} V^{\dagger} \to z U_{z}(0) V e^{igaL} V^{\dagger}
U^{\dagger}_{z}(0)
\label{19kca}
\end{equation}
under gauge transformations belonging to the homotopy class labeled
by the
center element $z$. Due to the restriction to the fundamental domain
it is again difficult
to directly read off the transformation properties of the $a^{p}$.
The presence of the
Jacobian allows us to proceed without detailed knowledge of this.

The preceding discussion refers to the Weyl gauge where we still
have the freedom of local, time independent gauge transformations.
The question arises
what happens to the gauge transformations after resolution of
Gauss's law,
and how they are connected to the formal symmetries of the gauge
fixed
Hamiltonian.

On general grounds, one would expect the following: Topologically
trivial gauge
transformations can be generated by the Gauss law operator and
should
be reduced to {\bf 1} in the physical space. All non-trivial
transformations
$U_{z}(x)$ belonging to a given $z$ are deformable into each other
and therefore equivalent in the physical
space. Thus after gauge fixing, we expect a global residual symmetry
with $Z_N$ group structure as the only remnant of the original local
gauge symmetry,
directly exhibiting the topology of the large gauge transformations.

How do the local gauge transformations $U(x)$ behave under
gauge fixing? In QED$_{4}$ or the Schwinger model, it is indeed easy
to
follow this explicitly \cite{LNOT1,LNOT2}. It was found
that the residual symmetry
(here: only a U(1) displacement symmetry) corresponds directly to
the homotopically
non-trivial gauge transformations.
In QCD, the situation is less clear \cite{LNT}.
One can show that any gauge transformation is reduced in the
process
of gauge fixing to a residual symmetry transformation, but the
connection
between the original gauge function and the particular residual
symmetry transformation
could not be established in all details due to technical difficulties.
Besides, the large number of formal symmetries of the type
(\ref{19h}) -- (\ref{19k})
does not seem to match the expected group structure ($Z$ for the
case of QCD$_{4}$, $Z_{N}$ for the case of QCD$_{2}^{\rm adj}$).
The present model will enable us to clarify this issue. Rather than
trying to
do it formally, we proceed to concrete applications to SU(2) and SU(3)
where the general theoretical
expectations will indeed be confirmed.

\section{The structure of the vacuum state}

Once the quantum-mechanical reduction of QCD$_{2}^{\rm adj}$ is
completed  we can proceed to determine the wave function of the
vacuum
state. The construction can be carried out in the Born-Oppenheimer
approximation -- we first freeze the variables $a^p$ and calculate the
effective
potential energy as a function of $a^p$ by ``integrating out the
fermion degrees
of freedom". Then we study quantum mechanics of the variables
$a^p$
living on the manifolds specified above. At this point the boundary
conditions on $\tilde\Phi$ -- vanishing of the ``radial"
wave function at zeroes of the Jacobian -- become important. The use
of the
Born-Oppenheimer approximation is justified {\em a posteriori} since
the  typical frequencies of the quantum mechanical variables
$a^p$
are of order $g$ while those characteristic of the fermion degrees of
freedom
are of order $1/L$, (almost) everywhere inside the fundamental
domain. If $gL\ll 1$, as we shall assume, the fermion frequencies
are much larger than those referring to $a^p$. The Born-
Oppenheimer
approximation and the effective quantum-mechanical description in
terms
of $a^p$ may fail only near exceptional points where the fermion
levels
cross the zero-energy mark. These points exist and play their role in
our analysis. They will be discussed in due course.

The general strategy in calculating the effective $a^p$-dependent
potential is
the same as the one usually applied in the Hamiltonian approach to
the Schwinger
model, see e.g. Ref. \cite{Shifman4}. For each given value of $a^p$ we
fill
the Dirac sea appropriately, so that all negative energy levels are
occupied
and all positive energy levels are free, and then find the energy of
the
Dirac
sea thus constructed. Since some of the fermion levels cross the
zero-energy mark at certain values of $a^p$ inside
the fundamental domain, the Dirac sea must be restructured
correspondingly. Hence, in different sectors of the fundamental
domain
the fermion component of the wave function is different, and the
effective potential energy for the $a^p$'s is also different. The full
vacuum
wave function is a linear combination of the wave functions in
different
sectors. We shall explicitly construct the fermion component in
each of the sectors, find the effective potential, and solve the
Schr\"{o}dinger equation in the variables $a^{p}$. Two specific cases
-- SU(2) and SU(3) -- will be considered separately.

\subsection{SU(2): Vacuum, symmetries and fermion condensate}

The physics content of the preceding formal developments is first
exhibited by
application to the technically simple case of SU(2) QCD$_{2}^{\rm
adj}$ on a small interval
($gL \ll  1 $). In this  limit, the relevant part of the Hamiltonian is
that of the
charged fermion, neutral gluon system of eqs. (\ref{10a}) and
(\ref{15})
\begin{equation}
\tilde{H}_{\varphi,a}=\int_{0}^{L} dx
\varphi^{\dagger}\sigma_{3}\left(\frac{1}{i}\partial_{1}-g a\right)
\varphi -
\frac{1}{2L}\frac{\partial^{2}}{\partial a^{2}}
\label{25}
\end{equation}
where we have denoted the fermion and gauge field degrees of
freedom by
\begin{equation}
\varphi=\varphi^{12} \quad , \quad  a=a^{3}=a_{1}-a_{2}
\label{26}
\end{equation}
to ease the notation.
This Hamiltonian acts on the space of ``radial'' wavefunctions
satisfying the
constraint
\begin{equation}
\tilde{\Phi}\left(a=2n\pi/gL\right)=0      \ .
\label{27}
\end{equation}
The neutrality condition (cf. eq. (\ref{20})) reads
\begin{equation}
Q^{3}|\Phi\rangle = \left(Q_{1}-Q_{2}\right)|\Phi\rangle =\int_{0}^{L}
dx
\varphi^{\dagger}\varphi |\Phi\rangle = 0 \ .
\label{28}
\end{equation}
The formal symmetry  transformations leaving the Hamiltonian
invariant (cf. eqs. (\ref{19h}) -- (\ref{19k})) can be
reduced here to
shifts of the gauge degrees of freedom accompanied by
phase changes of the charged fermions (central conjugations)
\begin{equation}
T_C(n): \qquad a \rightarrow  a+\frac{2\pi n}{gL}\quad , \quad
\varphi
\rightarrow  e^{2i\pi nx/L}\varphi
\label{29}
\end{equation}
and reflections accompanied by charge conjugations (interchange of
color labels 1,2)
\begin{equation}
R: \qquad a \rightarrow -a \quad , \quad\varphi \rightarrow
\varphi^{\dagger}
\label{30}
\end{equation}
(we ignore the global gauge transformations (\ref{19f}) which play
no role
in the physical sector; displacements $T_{D}(n,-n)$ on the other hand are
the same as $T_{C}(2n)$ for SU(2)).
We now proceed to determine in the small interval limit the ground
state of
this system. The normal mode expansion
\begin{equation}
\varphi\left(x\right)=\frac{1}{\sqrt{L}}\sum_{k=-
\infty}^{\infty}{a_{k} \choose b_{k}
}e^{ip_{k}x} \quad , \quad p_{k}=\frac{2
\pi}{L}\left(k+\frac{1}{2}\right)
\label{31}
\end{equation}
yields for the Hamiltonian $\tilde{H}_{\varphi,a}$  (eq. (\ref{25}))
and the
charge $Q^{3}$ (eq. (\ref{28}))
\begin{equation}
\tilde{H}_{\varphi,a}=\sum_{k=-
\infty}^{\infty}\left(a^{\dagger}_{k}a_{k}-
b^{\dagger}_{k}b_{k}\right)\left(p_{k}-ga\right)-
\frac{1}{2L}\frac{\partial^{2}}{\partial a^{2}}
\label{32}
\end{equation}
\begin{equation}
Q^{3}=\sum_{k=-
\infty}^{\infty}\left(a^{\dagger}_{k}a_{k}+b^{\dagger}_{k}b_{k}\right) \ .
\label{34}
\end{equation}
In the  adiabatic approximation, the ground state of the fast fermion
degrees of
freedom for a fixed value of the slow gauge degree of freedom is
obtained by
filling the fermion negative energy states. Assuming the $a$-modes
to be filled
for $k < M$ and the $b$-modes for $k\geq M^{\prime}$ the (heat
kernel
regularized) charge of such a state is given by
\begin{eqnarray}
Q^{3}|M,M^{\prime} \rangle & = & \lim_{\lambda \to 0}
\left(\sum_{k=-\infty}^{M-
1}e^{\lambda\left(p_{k}-ga\right)}+\sum_{k=M^{\prime}}^{\infty}e^{-
\lambda\left(p_{k}-ga\right)}\right)|M,M^{\prime} \rangle
\nonumber \\
& = & \left(M-
M^{\prime}\right)|M,M^{\prime} \rangle \ .
\label{35}
\end{eqnarray}
Neutrality (\ref{28}) requires the fermi levels to satisfy
\begin{equation}
M^{\prime}=M                  \ .
\label{36}
\end{equation}
The regularized energy of such a neutral state is
$$
\lim_{\lambda \to 0} \sum_{k=-
\infty}^{\infty}\left(a^{\dagger}_{k}a_{k}e^{\lambda\left(p_{k}-
ga\right)}-b^{\dagger}_{k}b_{k}e^{-\lambda\left(p_{k}-
ga\right)}\right)\left(p_{k}-ga\right)|M,M\rangle
=U_{M}\left(a\right)
|M,M\rangle\, ,
$$
\begin{equation}
U_{M}\left(a\right)=\frac{2\pi}{L}\left(M-
\frac{gLa}{2\pi}\right)^{2}\, .
\label{38}
\end{equation}
As is well known from similar treatments of the Schwinger model or
of  QCD
with fermions in the fundamental representation \cite{LTLY} the
adiabatic
ground state of the system is obtained by  adjusting  the occupation
of the
fermionic neutral states to the variations in the gauge degree of
freedom.
Obviously $U_{M}\left(a\right)$ is minimal with the following choice
of $M$,
\begin{equation}
\left|M\left(a\right)-\frac{gLa}{2\pi}\right|\leq \frac{1}{2}  \ .
\label{39}
\end{equation}
The occupation changes
whenever a pair of fermion levels crosses zero energy,
\begin{equation}
gLa=2\pi\left(n+\frac{1}{2}\right)\ .
\label{40}
\end{equation}
In this way, a periodic potential energy of the gauge degrees of
freedom
\begin{equation}
U\left(a\right):=U_{M\left(a\right)}\left(a\right)
\label{41}
\end{equation}
(where $M(a)$ is defined implicitly via eq. (\ref{39}))
is obtained (Fig. 5). The  analogy with the Schwinger model with
its discrete
translational symmetry generated by large gauge transformations
(topology of a circle)  is
only
superficial.
Physics consequences are very different due to the presence of the
constraint
(\ref{27}) on the radial wave function of QCD. This constraint
requires
solution of the Schr\"{o}dinger equation
in one fundamental interval defined by two consecutive zeroes of the
wave
function. Due to
the translational symmetry of $U$ we therefore can restrict the
calculations to
the fundamental domain discussed in subsect. 2.3,
\begin{equation}
0 \leq gLa \leq 2\pi
\label{42} \ .
\end{equation}
Then, the $k=0$ fermion $a$- and $b$-modes cross zero energy
in the middle, at
$gLa=\pi$.

As the characteristic property of QCD$_{2}$ with fermions in the
adjoint
representation we note the reflection symmetry of $U\left(a\right)$
around
$gLa=\pi$. The  double-well structure yields a two-fold degenerate
ground
state in the semiclassical limit with ground state wave functions
given by
\begin{equation}
\tilde{\Phi}_{I}\left(a\right)=
 \frac{2 (gL)^{3/4}}{\pi^{5/8}}\,
 a \, \exp\left\{
-\frac{gL}{2\sqrt{\pi}}a^{2}\right\} \ , \quad
\tilde{\Phi}_{II}\left(a\right)=\tilde{\Phi}_{I}\left(\frac{2\pi}{gL}-
a\right) \ ,
\label{43}
\end{equation}
corresponding to localization close to one of the two minima of the
potential
energy (see Fig. 5). As a  further consequence of  the Jacobian in the
kinetic energy
of $a$,
we note the vanishing of the wave function  at the minima of the
potential
energy. The ground state energy is
\begin{equation}
E_{0}=\frac{3}{2}\frac{g}{\sqrt{\pi}} \ .
\label{44}
\end{equation}

It is instructive to compare this result with the one obtained for
fermions in
the fundamental representation. The result for the potential energy
in the
corresponding adiabatic approximation is
\begin{equation}
U^{f}\left(a\right)=\frac{4\pi}{L}\left(M-\frac{gLa}{4\pi}\right)^{2}
\label{46}
\end{equation}
while the constraint (\ref{27}) remains unchanged.
The effective reduction in the coupling constant by a factor of 2
follows
directly from the difference in the minimal substitution which
couples
fermions and gauge degrees of freedom via $a_{1}$ and $a_{2}=-
a_{1}$ for
fundamental and via $a_{2}-a_{1}$ for adjoint fermions.
As a consequence of this reduced coupling $U^{f}\left(a\right)$
depends on $a$ monotonically  in the
fundamental interval and no degeneracy in the
ground
state occurs.

At this point the origin of the symmetry of the adiabatic potential
energy in
one fundamental interval remains to be understood.
The discrete translational symmetry of the potential energy
$U\left(a\right)$
of eq. (\ref{41})
\begin{equation}
U\left(a+\frac{2n\pi}{gL}\right)=U(a)
\label{47}
\end{equation}
 and the reflection symmetry
\begin{equation}
U(-a)=U(a)
\label{48}
\end{equation}
around $a=0$
are of no direct relevance, since they relate wave functions in
dynamically
decoupled intervals. These symmetries taken together however
entail
\begin{equation}
U\left(\frac{\pi}{gL}+a\right)=U\left(\frac{\pi}{gL}-
a\right) ,
\label{49}
\end{equation}
i.e., a reflection symmetry around the center of the fundamental
interval.
Although derived in the adiabatic approximation, it is
straightforward to
generalize this argument and to identify an exact and relevant
symmetry for
the system under consideration.
We start with the formal and in the above sense irrelevant
symmetries of eqs.
(\ref{29}), (\ref{30}). The following particular combination of a
central conjugation
and a reflection
\begin{equation}
S =T_C(1)R :  \quad \quad S a S^{\dagger} = -a+\frac{2\pi}{gL} \quad,
\quad
S \varphi S^{\dagger} = e^{2i\pi x/L} \varphi^{\dagger}
\label{50}
\end{equation}
connects states with the gauge degrees of freedom restricted to the
fundamental interval and therefore represents a relevant symmetry
of the system.
This operator can be chosen to satisfy
\begin{equation}
S^{2}= 1 \quad \mbox{and} \quad [S,H] = 0
\label{50a}
\end{equation}
and the stationary states can be classified generally as symmetric
and
antisymmetric ones.
$S$ transforms the wavefunctions $\tilde{\Phi}_{I}(a)$ and
$\tilde{\Phi}_{II}(a)$
into each other, cf. eq. (\ref{43}).
The action of $S$ on the fermionic states $|M,M\rangle $ (cf. eq.
(\ref{36})) can
easily be calculated,
\begin{equation}
S|M,M\rangle =e^{i\alpha_{M} } |-M+1,-M+1\rangle \ .
\label{52}
\end{equation}
Here, the phase $e^{i\alpha_M }$ is not determined by the above
properties of $S$, except for the condition
\begin{equation}
e^{i\alpha_M}= e^{-i \alpha_{-M+1}}
\label{52a}
\end{equation}
which follows from $S^{2}=1$.
Using only the fermionic part of $S$, this implies
the reflection symmetry of the adiabatic potential
\begin{equation}
\langle M,M|H_{\varphi}\left(a\right)|M,M\rangle =\langle -M+1,-
M+1|H_{\varphi}\left(-a+\frac{2\pi}{gL}\right)|-M+1,-M+1\rangle \ .
\label{53}
\end{equation}
For fermions in the fundamental representation, also both types of
formal
symmetries are present but the displacements in the gauge variable
$a$
occur in twice as large steps $a\rightarrow  a+\frac{4\pi}{gL}$. As a
consequence no relevant symmetry can be constructed  by
combining
displacements and reflections.

Finally we are able to write down the full vacuum state vector
which we take to be an eigenstate of the symmetry operator $S$
with eigenvalue $\pm 1$,
\begin{equation}
| \Psi_{\pm} \rangle = \frac{1}{\sqrt{2}} \left\{ |0,0\rangle
|\tilde{\Phi}_{I} \rangle
\pm e^{i\alpha_0 } |1,1\rangle |\tilde{\Phi}_{II} \rangle \right\} \, ,
\label{53a}
\end{equation}
where $\alpha_0$ is the phase introduced in eq. (\ref{52}).

Thus, we arrive at two vacua: one is described by $|\Psi_+\rangle$,
the other by $|\Psi_-\rangle$. The two linear conbinations in eq.
(\ref{53a})
are  the $Z_2$ analog of the $\theta$ vacua of QCD$_4$ where the
relevant symmetry is $Z$. The necessity of superimposing
the states $I$ and $II$ in the vacuum wave function, eq. (\ref{53a}),
is due to the fact that only under this choice the property of the
cluster decomposition will be satisfied. Moreover, if a fermion mass
term $m\bar\psi\psi$ is introduced as a small perturbation, physics
will be  smooth in $m$ under the choice (\ref{53a}). On the contrary,
had
we chosen the states $I$ and $II$ as the vacuum states, introducing a
small mass term would result in a drastic restructuring.

With these considerations we have succeeded in clarifying the
symmetry
properties of (SU(2)) QCD$_{2}^{\rm adj}$ in the canonical
formulation
with the unconstrained
degrees of freedom. The crucial element in the construction, valid
beyond
this particular model, is the emergence of relevant symmetry
transformations
by combining a sequence of irrelevant ones.

Using the above  expression for the vacuum it is straightforward to
calculate the fermion condensate
$\langle\bar\psi^{a}\psi^{a}\rangle$.
The contribution from the charged fermions to the condensate
operator
is
\begin{equation}
{\cal C}  =  \frac{1}{L} \int_{0}^{L} dx  \left( \varphi^{\dagger}
\gamma^0 \varphi
+ \varphi \gamma^{0} \varphi^{\dagger} \right)
 =   \frac{2i}{L} \sum_{k} \left( b_{k}^{\dagger} a_{k} -
a_{k}^{\dagger} b_{k} \right)
\label{53b}
\end{equation}
(the neutral fermions $\psi^{3}$ do not contribute in this
approximation).
This yields the condensate
\begin{equation}
\langle \Psi_{\pm} | \bar{\psi}^{a} \psi^{a} | \Psi_{\pm} \rangle
= \pm \frac{2}{L} \sin \alpha_0
\langle \tilde{\Phi}_{I}| \tilde{\Phi}_{II} \rangle \ .
\label{53c}
\end{equation}
with the overlap of the
two localized wave functions of the gauge degree of freedom given
by
\begin{equation}
\int_{0}^{2\pi/gL} da
\tilde{\Phi}_{I}\left(a\right)\tilde{\Phi}_{II}\left(a\right)=
\frac{4\pi^{3/2}}{gL}\exp \left(-
\frac{\pi^{3/2}}{gL}\right) .
\label{45}
\end{equation}

In addition to the expected  non-perturbative exponential
suppression ($gL\ll
1$) the overlap (\ref{45}) is actually enhanced by an inverse power
of $gL$ in the
prefactor. This is a direct consequence of the fact that
 $\tilde\Phi$ has nodes at the boundaries of
the fundamental domain. As a consequence of the constraint
(\ref{27})  on
the ``radial"
wave function, the system is pushed away from the minima of the
potential
energy into the classically forbidden region of the potential barrier.
The result in eq. (\ref{45})
agrees with the result obtained previously in Ref.
\cite{Smilga} by functional integration in the Euclidean formulation
of
QCD$_{2}^{\rm adj}$ as far as the exponent and the $gL$-dependence
of the prefactor is concerned.
Without the Jacobian, we would
have missed the factor
$(gL)^{-1} $.

The value of the condensate depends explicitly on the
choice
of phases. To clarify this dependence we observe that
in addition to the scalar condensate  one might also use
the pseudoscalar condensate  to characterize the ground state.
The value of this condensate is
\begin{equation}
\langle \Psi_{\pm} | \bar{\psi}^{a} \gamma^{5}
 \psi^{a} | \Psi_{\pm} \rangle
= \pm \frac{2i}{L} \cos \alpha_0
\langle \tilde{\Phi}_{I}| \tilde{\Phi}_{II} \rangle \ ,
\label{53d}
\end{equation}
which therefore can be combined with the scalar condensate to form
a complex
quantity with the modulus being independent of the phase $\alpha_0 $.
The arbitrariness in the phase can be removed by adding a weak
disturbance
to the Hamiltonian which will lift the twofold degeneracy
(eq. (\ref{53a})). Choosing
 this disturbance to be proportional to a scalar (mass) term
($ \propto \bar{\psi}^{a}  \psi^{a} $) or a pseudoscalar term
($ \propto \bar{\psi}^{a} \gamma^{5}  \psi^{a} $) fixes $\alpha_0 $ to be
$\pm \pi/2$ and $\pi$ respectively.

\subsection{Corrections to the adiabatic approximation}

We discuss here the corrections to the small interval limit. The
Coulomb energy (eq. (\ref{16})) acts as  a   centrifugal barrier on the
gauge degrees of freedom which in general are therefore prevented,
beyond the effect of the Jacobian, to approach the
 points $gLa=2n\pi$.
The relevant part of the Coulomb energy is determined by the
off-diagonal color charge densities which in turn contain both charged
and neutral fermion fields. In the adiabatic, small interval limit the
ground state of the neutral fermions is that of the
non-interacting system. The normal mode expansion of the hermitian
neutral field is
\begin{equation}
\psi^{3}\left(x\right)=\frac{1}{\sqrt{L}}\sum_{k\geq
0}\left[\left(c_{k}e^{ip_{k}x}+c^{\dagger}_{k}e^{-
ip_{k}x}\right){1\choose 0}+\left(d_{k}e^{ip_{k}x}+d^{\dagger}_{k}e^{-
ip_{k}x}\right){0\choose 1}\right] .
\label{54}
\end{equation}
The free field Hamiltonian (\ref{14}) reads in terms of these creation
and annihilation operators
\begin{equation}
H_{\psi}= \sum_{k\geq 0}p_{k}\left(c^{\dagger}_{k}c_{k}-
d^{\dagger}_{k}d_{k}\right)
\label{55}
\end{equation}
and therefore the ground state satisfies
\begin{equation}
c_{k}|0_{\psi}\rangle =0 \quad ,\quad
d^{\dagger}_{k}|0_{\psi}\rangle =0 \quad ,\quad k\geq 0 \ .
\label{56}
\end{equation}
With the choice (\ref{42}) of the fundamental interval, the
Coulomb energy becomes singular if, for $n=0,1$ or $-1$,  certain elements of
the matrix
\begin{equation}
\tilde{\rho}^{\,ij}\left(n\right)=\int_{0}^{L} dx e^{2i\pi n
x/L}\rho^{ij}\left(x\right)
\label{57}
\end{equation}
are non-vanishing when acting on the ground state. For the case of
SU(2) in particular,
the relevant part of the Coulomb energy is
\begin{equation}
\frac{g^{2}}{L} \sum_{n} \frac{\tilde{\rho}^{21}(n) \tilde{\rho}^{12}(-
n)+
\tilde{\rho}^{12}(-n) \tilde{\rho}^{21}(n)}{\left( 2\pi n /L - ga
\right)^{2}}
\label{57a}
\end{equation}
so that the potentially dangerous terms are $\tilde{\rho}^{12}(0),
\tilde{\rho}^{21}(0),
\tilde{\rho}^{12}(-1)$ and $\tilde{\rho}^{21}(1)$.

Using the normal mode expansions (eqs. (\ref{31}), (\ref{54})), the
off-diagonal charge
is given by
\begin{equation}
\tilde{\rho}^{12}\left(0\right)=\sum_{k \geq 0}\left(a_{-k-
1}c_{k}+a_{k}c_{k}^{\dagger}+b_{-k-
1}d_{k}+b_{k}d_{k}^{\dagger}\right) \ ,
\label{58}
\end{equation}
with $\tilde{\rho}^{21}(0) = \left[ \tilde{\rho}^{12}(0)
\right]^{\dagger} $.
Explicit calculation shows  the direct product of the neutral states,
$|M,M\rangle $ with $M=0$  (cf. eqs. (\ref{35}), (\ref{36})) and
$|0_{\psi} \rangle$ of eq. (\ref{56})
to be an eigenstate of the off-diagonal charge with vanishing eigenvalue
\begin{equation}
\tilde{\rho}^{12}\left(0\right)|0,0;0_{\psi}\rangle
=\tilde{\rho}^{21}\left(0\right)|0,0;0_{\psi}\rangle =0 \ ,
\label{59}
\end{equation}
i.e., the adiabatic ground state is a singlet state with respect to the
fermionic color charge. Under reflections  generated by $S$ the
off-diagonal charges transform as
\begin{eqnarray}
S\tilde{\rho}^{12}\left(0\right)S^{\dagger}& =& -\tilde{\rho}^{21
\dagger}\left(1\right)
\nonumber \\
S\tilde{\rho}^{21}\left(0\right)S^{\dagger}& =& -\tilde{\rho}^{12
\dagger}\left(-1\right)
\label{60}
\end{eqnarray}
which, together with the transformation properties of the states (cf.
eq. (\ref{52})) yields
\begin{equation}
\tilde{\rho}^{21}\left(1\right)|1,1;0_{\psi}\rangle
=\tilde{\rho}^{12}\left(-1\right)|1,1;0_{\psi}\rangle =0 \ .
\label{61}
\end{equation}
Thus in the evaluation of  the adiabatic ground-state expectation
value of the color electrostatic energy (\ref{16})
\begin{equation}
\theta\left(\pi-gLa\right)\langle
0,0;0_{\psi}|H_{C}|0,0;0_{\psi}\rangle +\theta\left(gLa-\pi
\right)\langle 1,1;0_{\psi}|H_{C}|1,1;0_{\psi}\rangle
\label{62}
\end{equation}
the residues of the singularities in the Laurent expansion in $gLa$
are zero and the contribution to the energy is  $\propto g^{2}L$ and
therefore negligible in the small interval limit.
For finite interval length, vanishing  fermionic color charges are not
anymore energetically favoured and as a  result of the  electrostatic
centrifugal barrier, (radial) wave function components will be
admixed which vanish faster than required by the
Jacobian (this effect has been demonstrated to actually occur for
static
color charges in the fundamental representation \cite{ES}).

\subsection{The vacuum in the SU(3) case}

The calculation of the vacuum structure in the adiabatic
approximation described
above for the SU(2) case can be generalized to SU(N). We start from
the $\frac{1}{2}N(N-1)$ charged fermion fields $\varphi^{ij}
(i<j)$ introduced in eq. (\ref{11})
and the normal mode expansion analogous to eq. (\ref{31}),
\begin{equation}
\varphi^{ij}\left(x\right)=\frac{1}{\sqrt{L}}\sum_{k=-
\infty}^{\infty}{a_{k}^{ij} \choose b_{k}^{ij}
}e^{ip_{k}x} \quad , \quad p_{k}=\frac{2
\pi}{L}\left(k+\frac{1}{2}\right) \ .
\label{63}
\end{equation}
The adiabatic fermion ground state is assumed to have
the ``$a^{ij}$-modes"
filled for $k < M_{ij}$
and the ``$b^{ij}$-modes" filled for $k \geq M_{ij}'$. These states are
eigenstates of
$H_{\varphi}(a)$, eq. (\ref{15}), with regularized eigenvalues
\begin{equation}
H_{\varphi}\left(a\right)|\vec{M},\vec{M}^{\prime}\rangle =
\frac{\pi}{L}\sum_{i<j}\left[\left(M_{ij}-\frac{gL}{2\pi} (a_{i}-
a_{j})\right)^{2}+\left(M_{ij}^{\prime}-\frac{gL}{2\pi} (a_{i}-
a_{j})\right)^{2}\,\right]
|\vec{M}, \vec{M}' \rangle
\label{65}
\end{equation}
($\vec{M}, \vec{M}'$ denote $N(N-1)/2$ dimensional vectors with
components
$M_{ij}, M_{ij}', i<j$).
For given values of the gauge variables $a_{i}$ the choice
\begin{equation}
M_{ij}=M_{ij}^{\prime} \quad ,\quad \left|M_{ij}\left(a\right)-
\frac{gL}{2\pi}(a_{i}-a_{j})\right|\leq \frac{1}{2} \quad , \quad i<j
\label{66}
\end{equation}
yields the lowest eigenvalue, i.e., the effective potential.
Of course, we have to verify that these states satisfy the neutrality
condition (\ref{20}). Exactly like in the SU(2) case, one first
shows that (after regularization)
\begin{equation}
\int_{0}^{L} dx \varphi^{ij\dagger} \varphi^{ij} |\vec{M},
\vec{M}^{\prime} \rangle =
\left( M_{ij} - M_{ij}' \right) | \vec{M},\vec{M}^{\prime} \rangle \ ,
\label{66b}
\end{equation}
where the indices $(i,j)$ are kept fixed. For the state of lowest
energy, the right
hand side vanishes. Hence
in the expression (\ref{21}) for the neutral charge operators $Q_{i}$,
each term in the sum vanishes separately when acting
on the adiabatic ground state, so that this latter is indeed a physical
state.

Although this part of the calculation can trivially be done for
arbitrary
$N$, the following step, namely solving the Schr\"{o}dinger equation
in the gauge variables and discussing in detail the symmetries,
condensates,
spectral flow etc. quickly becomes complicated due to the increasing
number
of variables and the geometrical intricacies associated with the
fundamental domain. For our purpose it is more useful to exhibit in
detail
the results for SU(3). Together with the SU(2) case, this provides
enough understanding to be able to foresee what will happen for
larger $N$,
at least qualitatively.

For the  explicit calculation in SU(3), it is preferable to change from
the diagonal matrix elements $a_{i}$ of the gauge degrees of freedom
to the neutral amplitudes $a^{3}, a^{8}$ (cf. eq. (\ref{19f})),
which are the natural variables for the kinetic energy term of the
gauge field
and, more importantly, the independent ones.
The Schr\"{o}dinger equation in the ``radial" form now corresponds
to a
quantum mechanical problem on a plane,
\begin{equation}
\left[- \frac{1}{2L} \left( \frac{\partial^{2}}{\partial (a^{3})^{2}} +
 \frac{\partial^{2}}{\partial (a^{8})^{2}}  \right) + U_{eff}(a^{3}, a^{8})
\right] \tilde{\Phi}(a^{3}, a^{8})
= E \tilde{\Phi}(a^{3}, a^{8})
\label{66c}
\end{equation}
with the condition that $\tilde{\Phi}$ vanishes along the boundary
of the fundamental triangle (cf. Fig. 4 and eq. (\ref{19ga})).
The effective potential is given by
\begin{equation}
U_{eff}\left(a^{3},a^{8}\right)=U\left(a^{3}\right)+
U\left(\frac{1}{2}(a^{3}+\sqrt{3}a^{8})\right)+U\left(\frac{1}{2}(-
a^{3}+\sqrt{3}a^{8})\right)
\label{67}
\end{equation}
with $U(a)$ defined in eqs. (\ref{39}) -- (\ref{41}).
As indicated in Fig. 6, we can subdivide further the fundamental
domain
into 4 congruent triangles, in each of which the adiabatic potential is
that of
an isotropic, 2-dimensional harmonic oscillator,
\begin{equation}
U^{I}\left(a^{3},a^{8}\right)
=\frac{3 g^{2}L}{4\pi}\left((a^{3})^{2}+(a^{8})^{2}\right)
\label{69}
\end{equation}
\begin{eqnarray}
U^{II}\left(a^{3},a^{8}\right)& = & U^{I}\left(a^{3}-
\frac{2\pi}{gL},a^{8}-\frac{2\pi}{\sqrt{3} gL}\right)\nonumber \\
U^{III}\left(a^{3},a^{8}\right)& = & U^{I}\left(a^{3}-
\frac{2\pi}{gL},a^{8}+\frac{2\pi}{\sqrt{3} gL}\right)\nonumber \\
U^{IV}\left(a^{3},a^{8}\right)& = & U^{I}\left(a^{3}-\frac{4\pi}{3
gL},a^{8}\right)+\frac{2\pi}{3L} .
\label{70}
\end{eqnarray}
Generalizing the reflection symmetry of the SU(2) case, this potential
also
exhibits discrete symmetries: It is invariant under rotations around
the center of
the fundamental triangle by 120$^\circ$ or 240$^{\circ}$, as well as
under reflections with respect to 3 lines joining its center with each
corner (see Figs. 7 and 8).

As in the SU(2) case, the discrete rotational symmetry is a property
of the exact theory, valid
beyond the
adiabatic approximation. We can
follow the analysis of the SU(2) case.
As emphasized above, the formal residual symmetries are separately
irrelevant, since
they do not respect the fundamental domain. Guided by the
observation
that we are dealing with a discrete subgroup of the 2-dimensional
Euclidean
group E$_{2}$ in the $a^{3}, a^{8}$ plane, it is easy to identify the
particular combination of the symmetries
which does not lead out of the fundamental domain: The cyclic permutation
\begin{equation}
C: \quad \quad  \left( \begin{array}{ccc} 1 & 2 & 3 \\ 3 & 1 & 2
\end{array} \right)
\label{70a}
\end{equation}
of the color labels rotates the triangle by 120$^\circ$ around the
origin of the $(a^{3},a^{8})$ plane into another
fundamental domain (see Fig. 4). It can be shifted back into the
original domain
by combining a displacement ($k_{1}=0, k_{2}=-1, k_{3}=1$) with a
central conjugation ($n=1$),
\begin{equation}
S = T_{D}(0,-1,1)T_{C}(1) C \ .
\label{70b}
\end{equation}
The operator $S$ acts as follows on our variables,

\begin{eqnarray}
 Sa^{3}S^{\dagger} & = & a^{3}\,' =  -
\frac{1}{2}\left(a^{3}+\sqrt{3}a^{8}\right)+ \frac{2\pi}{gL}\nonumber
\\  Sa^{8}S^{\dagger}& = & a^{8}\,' = \frac{1}{2}\left(\sqrt{3}a^{3}-
a^{8}\right)- \frac{2\pi}{\sqrt{3}gL}\label{71} \\
S\varphi^{12}S^{\dagger} & = & e^{i 2\pi x/L}\varphi^{13\dagger}
\quad , \quad S\varphi^{23}S^{\dagger}=e^{-i2\pi x/L}\varphi^{12}
\quad , \quad S\varphi^{13}S^{\dagger}=\varphi^{23\dagger} \ .
\nonumber
\label{72a}
\end{eqnarray}
$S^{2}$ is the inverse operator of $S$, so that $S^{3}= 1$. For later
convenience we also
note the inverse relations
\begin{eqnarray}
 S^{\dagger}a^{3}S & = &  a^{3}\,'' =  - \frac{1}{2}\left(a^{3}-
\sqrt{3}a^{8}\right)+ \frac{2\pi}{gL} \ , \nonumber \\
S^{\dagger}a^{8}S& =&  a^{8}\,'' =
-\frac{1}{2}\left(\sqrt{3}a^{3}+a^{8}\right)+ \frac{2\pi}{\sqrt{3}gL} \ .
\label{72}
\end{eqnarray}
This is the quantum mechanical formulation of the $Z_{3}$ symmetry
of rotations of the fundamental triangle.
Since any residual symmetry transformation can be written as a
permutation followed by a shift,
it is easy to see that this exhausts the exact symmetries of the model.
The additional reflection
symmetries exhibited by the effective potential are not symmetries
of the full Hamiltonian.
Geometrically, the reason is the following: These
reflections
involve non-cyclic permutations of the color indices. Such
permutations change the orientation of the basic triangle in the plane
by $\pm$ 60$^\circ$,
so that it cannot be mapped back onto the fundamental domain by a
subsequent
translation.

Let us now again build the vacuum state vector. First consider the
fermions.
The occupation of the Dirac sea changes across the boundary of the
internal
triangle (IV) in Fig. 6 where pairs of fermion levels cross zero
energy, one pair at each boundary.
We denote the adiabatic fermionic states by $|M_{12}, M_{13},
M_{23} \rangle $
(the $M_{ij}'$ are redundant owing to the neutrality condition). In Fig.
6,
we also display the ``fermi levels" $M_{ij}$ in the four different
sectors of the fundamental domain.
The action of $S$ on these states in general is
\begin{equation}
S|M_{12}, M_{13}, M_{23} \rangle = |M_{23}+1, -M_{12}+1, -M_{13}
\rangle \ ,
\label{74}
\end{equation}
where we have made a definite choice of the phases.
Specifically, for the case at hand,
\begin{displaymath}
S|0,0,0 \rangle = |1,1, 0\rangle \ , \quad S|1,1, 0 \rangle = S|1,0,-1
\rangle \ ,
\quad S |1,0,-1 \rangle = |0,0,0 \rangle \ ,
\end{displaymath}
\begin{equation}
 S|1,0,0 \rangle = |1,0,0 \rangle   \ ,
\label{75}
\end{equation}
in accordance with the discrete rotational symmetry. The
Schr\"{o}dinger
equation for $\tilde{\Phi}$ has clearly a 3-fold degeneracy with
wavefunctions localized near the three corners related by actions of
$S$,
\begin{eqnarray}
\tilde{\Phi}_{II}(a^{3}, a^{8}) & = & S \tilde{\Phi}_{I}(a^{3},a^{8})
=\tilde{\Phi}_{I}
\left(a^{3}\,'',a^{8}\,'' \right) \ , \nonumber \\
\tilde{\Phi}_{III}(a^{3}, a^{8}) & = & S \tilde{\Phi}_{II}(a^{3},a^{8}) =
\tilde{\Phi}_{I}\left(a^{3}\,', a^{8}\,' \right) \ .
\label{76}
\end{eqnarray}
with the primed and double-primed arguments defined in eqs.
(\ref{71}), (\ref{72}).
The ground state wavefunction in particular corresponds to an $n=3$,
$f$-wave
excited state of the 2-dimensional isotropic harmonic oscillator,
due to the boundary condition that it has two nodal lines subtending
an
angle of 60$^\circ$,
\begin{equation}
\tilde{\Phi}_{I}(a^{3}, a^{8}) = \frac{2}{gL} \left( \frac{32
\nu^{4}}{\pi} \right)^{1/2}
\rho^{3} \cos (3 \phi) e^{- \nu \rho^{2}}
\label{77}
\end{equation}
with plane polar coordinates defined as
\begin{equation}
\frac{gLa^{3}}{2} = \rho \cos \phi \ , \quad \frac{gLa^{8}}{2} =
\rho
\sin \phi
\label{78}
\end{equation}
and
\begin{equation}
E_{0} = 4 \omega \ , \quad \omega^{2} = \frac{3g^{2}}{2\pi} \ , \quad
\nu = \frac{1}{gL} \sqrt{\frac{6}{\pi}} \ .
\label{79}
\end{equation}
It is then easy to construct, with the above choice of
phases, simultaneous eigenstates of $H$ and $S$,
\begin{equation}
|\Psi_{z} \rangle = \frac{1}{\sqrt{3}} \left\{ |0,0,0 \rangle
|\tilde{\Phi}_{I} \rangle
+ z^{2} |1,1, 0 \rangle |\tilde{\Phi}_{II} \rangle
+ z |1,0,-1 \rangle |\tilde{\Phi}_{III} \rangle   \right\}
\label{80}
\end{equation}
with
\begin{equation}
S |\Psi_{z} \rangle = z |\Psi_{z} \rangle \ , \quad z \in \left\{1,
e^{2\pi i/3}, e^{4\pi i/3} \right\} \ .
\label{81}
\end{equation}
Since the fermionic components differ in the occupation of 2 pairs of
levels,
the bilinear condensate vanishes in this approximation.
On the other hand, one could construct an operator of 4th-order in
the
fermion fields with non-zero matrix elements between
the adiabatic fermion states in the sectors $I, II$ and $III$, but
vanishing contribution within one sector. In this case, a quartic
condensate,
$$
i\,\mbox{tr}\,\{\bar\psi (1-\gamma_5)\psi
\bar\psi (1-\gamma_5)\psi\}\,\,\,\mbox{or}\,\,\,
i\,\mbox{tr}\,\{\bar\psi (1+\gamma_5)\psi
\bar\psi (1+\gamma_5)\psi\}\,\,  ,
$$
appears being determined by the overlap of the bosonic
wavefunctions. The
relevant (bosonic) matrix element is again calculable
(for $gL \ll 1$) and yields
\begin{equation}
\langle \tilde{\Phi}_{I} | \tilde{\Phi}_{II} \rangle  = 2 \pi^{4} \nu^{2}
e^{-2\nu \pi^{2}/3}
\label{82}
\end{equation}
The effect discussed above for SU(2) -- enhancement of the overlap
due to
the Jacobian -- is even more pronounced in SU(3) where it forces us into
higher harmonic oscillator states; it yields a prefactor $\sim (gL)^{-
2}$.

\subsection{Anomalies and  spectral flow of fermion levels}

Up to this point, the charge operators relevant for constructing the
vacuum state vectors
were those which enter the neutrality condition, cf. eqs.
(\ref{20}), (\ref{21}).
Let us now consider the corresponding {\em axial} charges,
\begin{equation}
Q_{5i}=
\sum_{k>i}(Q_{5})_{ik}-\sum_{k<i}(Q_{5})_{ki}
\label{82a}
\end{equation}
with
\begin{equation}
(Q_{5})_{ik}=\frac{1}{2}\int_{0}^{L} dx
\varphi^{ik \dagger}\gamma^5 \varphi^{ik} .
\label{ad2}
\end{equation}
They provide the most direct access to the issue of axial anomalies
in our framework. Exactly like the ordinary charges $Q_{i}$, the axial charges
$Q_{5i}$ will be regularized in a way which is invariant under the
displacements and central conjugations (cf. eqs. (\ref{19h}), (\ref{19i})).
Heat kernel regularization yields
\begin{equation}
(Q_{5})_{ij}|\vec{M},\vec{M} \rangle
= \left(  M_{ij}- \frac{gL}{2\pi} \left(a_{i}-a_{j}\right)
\right)|\vec{M},\vec{M} \rangle \ .
\label{82c}
\end{equation}
The presence of $\gamma^5$ ($=\sigma_{3}$) introduces a minus sign between
the contributions from right- and left-handed fermions, so that two terms
which cancel in the ordinary charge (cf. eqs. (\ref{28}), (\ref{66b})) add
up in the axial
charge, leading to a dependence of $Q_{5i}$ on the gauge variables $a_{i}$.
It is convenient to define  purely fermionic operators by
\begin{equation}
(\tilde{Q}_{5})_{ij}=(Q_{5})_{ij} + \frac{gL}{2\pi}(a_{i}-a_{j})
\label{82e}
\end{equation}
and the corresonding combinations ($\tilde{Q}_{5i}$, $\tilde{Q}_{5}^p$, cf. eq.
(\ref{82a})) which by construction are not invariant under displacements
or central conjugations. However the charges $\tilde{Q}_{5}^p$ are
 conserved in the small interval limit. On the other hand, the invariant
axial charges $Q_{5}^p$ do not commute with the
small interval Hamiltonian (cf. eq. (\ref{13})), but
yield the ``anomalous" result
\begin{equation}
\label{82f}
\left[K_{a}+H_{\rm ferm}, Q_{5}^p \right] = i \frac{Ng}{4\pi} \left(
e^p + e^{p \dagger} \right) \ .
\end{equation}
We note in passing the explicit expressions for the relevant
axial charges in SU(2) and SU(3),
\begin{equation}
{\mbox SU(2)}:\quad
Q_{5}^3 =\int_{0}^L dx \varphi^{12\dagger}\gamma^5 \varphi^{12}\ ,
\label{su2-cur}
\end{equation}
$$
{\mbox SU(3)}:\quad
Q_{5}^3 =\int_{0}^L dx \left( \varphi^{12\dagger}\gamma^5 \varphi^{12}
+\frac{1}{2}\varphi^{13\dagger}\gamma^5 \varphi^{13}
-\frac{1}{2}
\varphi^{23\dagger}\gamma^5 \varphi^{23}\ \right) \ ,
$$
\begin{equation}
Q_{5}^8 = \int_{0}^L dx \frac{\sqrt{3}}{2}\left(
\varphi^{13\dagger}\gamma^5 \varphi^{13}
+
\varphi^{23\dagger}\gamma^5 \varphi^{23}
\right)\ .
\label{su3-cur}
\end{equation}

At this point a comparison with the Schwinger model reveals the essential
structures underlying these definitions. In the Schwinger model, the
residual gauge symmetry is that of the displacements (similar to  (\ref{19h})).
Therefore, the above regularization renders both the corresponding vector
charge $Q$ and axial charge $Q_{5}$ gauge invariant -- i.e., invariant under
the residual gauge symmetry transformations. In SU(2) QCD, the residual
symmetry transformation $S$ of eq. (\ref{50}) consists of a combined reflection
and central conjugation.
Neither of the two axial charges $Q^{3}_{5}$, $\tilde{Q}_{5}^3$
(cf. eqs. (\ref{82c}),
(\ref{82e})) are invariant (nor is the charge $Q^3$, cf. eqs.
(\ref{34}), (\ref{35})).
The above regularization however guarantees that $Q^{3}_{5}$ transforms
covariantly under the residual gauge transformation
\begin{equation}
SQ_{5}^3S^{\dagger}=-Q_{5}^3 .
\label{ad1}
\end{equation}
The axial
charge $\tilde{Q}_{5}^{3}$ on the other hand transforms inhomogeneously
\begin{equation}
S\tilde{Q}_{5}^3S^{\dagger}=-\tilde{Q}_{5}^3+2  \, .
\label{ad13}
\end{equation}
For SU(3) similar arguments apply and corresponding relations can
be derived easily on the basis of the transfomation properties
\begin{eqnarray}
S(Q_{5})_{12}S^{\dagger}&=&-(Q_{5})_{13} \quad ,\quad
S(\tilde{Q}_{5})_{12}S^{\dagger}=-(\tilde{Q}_{5})_{13} + 1 \nonumber\\
S(Q_{5})_{13}S^{\dagger}&=&-(Q_{5})_{23} \quad ,\quad
S(\tilde{Q}_{5})_{13}S^{\dagger}=-(\tilde{Q}_{5})_{23} \nonumber\\
S(Q_{5})_{23}S^{\dagger}&=&\,(Q_{5})_{12} \ \ \quad ,\quad
S(\tilde{Q}_{5})_{23}S^{\dagger}=(\tilde{Q}_{5})_{12} - 1
\label{ad3}
\end{eqnarray}

The gauge variant charges $\tilde{Q}_{5}^{p}$ commute with the
small interval Hamiltonian ($K_{a}+H_{\rm ferm}$) of the system.
Therefore, in the small interval limit, all eigenstates can be
classified with respect
to the values of the ${\tilde Q}_5^p$ charges. Moreover, in the SU(3) theory
${\tilde Q}_5^3$ and ${\tilde
Q}^8_5$
commute with each other.
The states corresponding to different eigenvalues of ${\tilde Q}_5^p$
are necessarily orthogonal to each other.

It is also possible to formulate in the small interval limit
the axial anomaly in a ``local" form,
\begin{equation}
\partial^\mu J_{\mu 5}^p =\frac{Ng}{4\pi L} \left(e^p+e^{p\, \dagger}
\right)\ ,
\label{anomaly}
\end{equation}
with the neutral components of the axial current
\begin{equation}
J_{\mu 5}^p = \frac{1}{2} \psi \gamma_{0} \gamma_{\mu} \gamma^5 T^p \psi \ .
\label{82g}
\end{equation}
This should be compared to the non-anomalous vector current $J_{\mu}^p=
\frac{1}{2} \psi
\gamma_{0}\gamma_{\mu} T^p \psi$, the neutral components of which are
conserved in the ordinary sense,
\begin{equation}
\partial^{\mu} J_{\mu}^p = 0 \ ,
\label{82ga}
\end{equation}
since in our specific gauge the only remnants of the gauge fields are the
neutral variables $a^p$.

In the canonical framework,  eq. (\ref{anomaly}) is obtained by
evaluating  the commutator
of the axial charge density with the small interval Hamiltonian.
The relevant Hamiltonian $H_{\varphi}(a)$ of eq. (\ref{15}) can be
equally well interpreted as the weak-coupling Hamiltonian of
$N(N-1)/2$ charged
particles coupled to $N-1$ different photons. Therefore in the
gauge fixed formulation and the weak coupling limit,  the non-abelian anomaly
of SU(N) QCD$_2$
is identical to the anomaly of an abelian U(1)$^{N-1}$ gauge theory.

By and large, the analysis of the anomaly of the Schwinger model can be
repeated
with minor modifications to establish in the weak coupling limit
index theorems, to classify the wave functions in different
sectors, and to address the issue of the spectral flow.

Let us first discuss this last point. As usual the
problem is
formulated as follows. One adiabatically varies the gauge field along
a certain
trajectory starting at $\{ a^p\}_i$ at $t=-T$ and arriving at $\{
a^p\}_f$
at $t=T$ (assuming $T\rightarrow\infty$). At each given value of $t$,
i.e., for a given gauge field, the energy eigenvalues of relevant
fermion
degrees of freedom in the first-quantized eigenvalue problem are
found. In
our case this is a particular simple excercise since the Coulomb
interaction
has been   neglected, the background gauge field configuration reduces
to spatial constants $a^p$, and then each fermion mode is the
eigenfunction
of the one-particle equation
$$
\sigma_3 \{(1/i)\partial_1 -g (a_i- a_j )\} \varphi_k^{ij} = E_k
\varphi_k^{ij}\, .
$$
where $k$ numbers the fermion modes.

At the next stage one studies evolution of $E_k$ versus $t$.  If some
of the
levels cross zero this phenomenon is in one-to-one correspondence
with the
anomalous non-conservation of some of the fermion charges, see e.g.
\cite{Shifman4}.  Moreover, using arguments similar to those of Ref.
\cite{Witten2} one can relate the fact of the crossover to the
occurence of the
zero modes in the two-dimensional Dirac operator in  background
fields
interpolating between $\{ a^p\}_i$ and $\{ a^p\}_f$.

Since the SU(2) and SU(3) theories differ in important technical
details it is
convenient to consider them separately. In the SU(2) theory we have
only
one gauge-field variable, $a^3$, and hence one topological charge
\begin{equation}
q = \frac{2g}{8\pi}\int \epsilon^{\mu\nu}F_{\mu\nu}^3 d^2 x
=\frac{gL}{2\pi}\left\{ a^3 (+T) - a^3 (-T)\right\} \, .
\label{su2-top}
\end{equation}
The interval of variation of the variable $a^3$ to be considered is the
fundamental interval $[0, 2\pi /(gL)]$. Two end-points of this
interval are
gauge-equivalent. The gauge transformation glueing the end-points
is a large
gauge transformation, see eq. (\ref{su2u}). Any trajectory connecting
them is
characterized by
$|q| =1$.  Eq. (\ref{anomaly}) shows then that in this transition
$$
\Delta Q_5^3 = 2\, ,
$$
i.e., two fermion levels -- one right-handed and one
left-handed -- cross the zero in the opposite directions when we
adiabatically proceed from the origin to $a^3 =\frac{2\pi}{gL}$. In agreement
 with our general remarks of above it does not come as a surprise
that in this  weak coupling limit the spectral flow of SU(2)
 QCD is identical to that
in the Schwinger model with one fermion field, see Fig. 1 in
\cite{Shifman4},
restricted to the interval $0<a<\frac{2\pi}{gL}$.

Proceeding to the SU(3) case we observe that the fundamental
domain
where the spectral flow is to be analysed is given by the triangle of
Fig. 6.
The three vertices of the triangle are gauge equivalent. The
corresponding large
gauge transformations glueing them are $U=\exp
\left\{\pm (2\pi i  x/L\sqrt{3})t^8 + (2\pi i x/L)t^3\right\}$ (we
remind that
$t^a=\lambda^a/2$ in the SU(3) theory). Although one can
consider any trajectory running inside the fundamental domain and
the
associated
spectral flow, of special interest are the trajectories starting at one
vertex
at $t=-T$ and ending up at another at $t=+T$.

Now we can introduce three topological charges
$$
q_\pm = \frac{3g}{16\pi}\int \, \varepsilon^{\mu\nu}
\left( F_{\mu\nu}^3\pm \frac{1}{\sqrt{3}}F_{\mu\nu}^8\right) d^2
x\, ,
$$
\begin{equation}
q = \frac{\sqrt{3}g}{8\pi}\int \, \varepsilon^{\mu\nu}
F_{\mu\nu}^8\,  d^2 x
\label{su3-top}
\end{equation}
subject to a constraint
$$
q_+ - q_- = q \, .
$$
On each of the three trajectories connecting different vertices the
absolute
value of one of the topological charges is 1 while for the two others we
have
1/2. The one with the maximal absolute value is relevant for the
given
trajectory.

For instance, if the trajectory runs along the upper side of the
triangle $a^8 = a^3/\sqrt{3}$, and the effective color current coupled
to the ``gluon field" is proportional to
$J_\mu^3 + J_\mu^8/\sqrt{3}$; from eq. (\ref{su3-cur}) we see that
the corresponding axial  charge is that of the two-flavor Schwinger
model ($\varphi^{12}$ and $\varphi^{13}$). The
spectral flow along this trajectory coincides with that of the
two-flavor
Schwinger model, with two pairs of levels crossing the zero-energy
mark.
The trajectories running along two other sides of the fundamental
triangle have the same properties modulo cyclic permutations of
$\varphi^{12}, \,\, \varphi^{13}$ and $\varphi^{23}$.

Generically, with the antiperiodic boundary conditions on the
fermion fields,
the lines where a pair of the fermion levels crosses zero form an
equilateral
triangle inside the fundamental domain, see Fig. 6.

In Ref. \cite{Smilga} the instanton solutions were found in
QCD$_2^{\rm adj}$
--
trajectories interpolating between the vertices of the fundamental
triangle. It
was shown that these solutions are accompanied by $2(N-1)$
fermion zero
modes for SU(N). Usually the fermion zero modes are associated with
index theorems of the Atiyah-Singer type. The latter relate the
difference in
the numbers of the left-handed and right-handed zero modes to the
topological charge of the given background field. There is obviously a
close
kinship between the index theorems, anomaly relations and the issue
of the
topological charge. In QCD$_4$ the index theorem was established in
Ref. \cite{Schwarz}, see also \cite{Coleman} for a thorough discussion.
We understand now that analogous index theorems exist in
QCD$_2^{\rm adj}$ but they necessarily involve a topological
charge whose definition is given in a particular gauge. For instance,
in the SU(2) theory
\begin{equation}
n_L-n_R = q\, , \quad n^+_R -n^+_L = q
\label{index}
\end{equation}
where $n_{L,R}$ and $n^+_{L,R}$ are the numbers of the
corresponding zero modes for $\psi$ and $\psi^+$, respectively.
Eq. (\ref{index}) explains why for any field trajectory interpolating
between $a^3=0$ and $a^3=\frac{2\pi}{gL}$, not necessarily the instanton
solution, the zero modes will persist.

\subsection{Symmetries beyond weak coupling}

So far our analysis has been performed in the small interval or weak
coupling limit. For the case of the Schwinger model or the U(1)
anomaly in QCD$_{4}$, corrections to the weak coupling limit
 do not affect the anomaly. This is different in the case of QCD$_2^{\rm adj}$.
The symmetry of the weak coupling limit is significantly higher than that
of the full theory. The full theory does not exhibit any continuous
axial symmetry. Before gauge fixing,  the only continuous symmetries are
gauge symmetries. After gauge fixing no continuous symmetry survives; the
axial symmetries of the weak coupling limit are manifestly broken by the
Coulomb interaction $H_{C}$ (cf. eqs. (\ref{13}), (\ref{16})). Beyond the weak
 coupling limit, the theory still exhibits a discrete axial symmetry.
The full Hamiltonian (\ref{13}) remains invariant under changes of sign
of either all the right- (upper components) or all the left-handed
(lower components) fermion fields. The unitary transformation, generating
sign changes of right-handed fields
\begin{equation}
\tilde{R}=\exp i\pi\left[\sum _{i,k(i<k)}\left(
(\tilde{Q}_{5})_{ik}+Q_{ik}\right)+\sum_{p=1}^{N-1}
\sum_{k\ge 0}c_{k}^{p\,\dagger}c_{k}^{p}\right]
\label{ad4}
\end{equation}
is a discrete symmetry transformation of QCD$_2^{\rm adj}$ beyond
the weak coupling limit (the vector charges $Q_{ik}$ are defined in analogy
to eq. (\ref{ad2})),
\begin{equation}
\tilde{R}H\tilde{R}^{\dagger}=H
\label{ad7}
\end{equation}
We can define similarly left-handed transformations. No new structures
however are encountered, since the transformation reversing the sign
of all fermion fields commutes with all other symmetry transformations.
Here it is not convenient to define a corresponding transformation
generated by the axial charge only. This would involve a redefinition of the
phase of all fermion fields which have been assumed to be real. We nevertheless
will refer to $\tilde{R}$  as a discrete  axial charge transformation.

The operators $c_{k}^{p\dagger},d_{k}^{p\,\dagger}$
 create neutral fermions  and are defined in analogy with the
SU(2) operators of eq. (\ref{54}). To characterize how this symmetry
is realized, we introduce the condensate operators
\begin{equation}
\Gamma_{0,\left(5\right)}=\sum _{i,k(i<k)} \varphi^{ik\dagger}\gamma^{0}\left(
\gamma^{5}\right)\varphi^{ik}+\sum_{p=1}^{N-1} \psi^{p}\gamma^{0}\left(
\gamma^{5}\right)\psi^{p}\ ,
\label{ad5}
\end{equation}
and it is easy to verify that these condensate operators are odd under the
discrete axial transformations
\begin{equation}
\tilde{R}\Gamma_{0,\left(5\right)}\tilde{R}^{\dagger}=
-\Gamma_{0,\left(5\right)}\ .
\label{ad6}
\end{equation}
A non vanishing scalar or pseudoscalar condensate associated
with these bilinear fermion operators can therefore develop
only if the system is not in an eigenstate of $\tilde{R}$,
\begin{equation}
\langle \Phi|\Gamma_{0,\left(5\right)}|\Phi \rangle \neq 0 \quad \mbox{only if}
\quad \tilde{R}|\Phi\rangle \neq \pm |\Phi\rangle .
\label{ad8}
\end{equation}
The appearence of quartic condensates on the other hand,
as considered above in connection
with SU(3), is not constrained by these symmetry properties. We obviously
have
\begin{equation}
\tilde{R} i \mbox{tr} \left\{\bar{\psi}(1\pm\gamma_{5})\psi
\bar{\psi}(1\pm\gamma_{5})\psi\right\}\tilde{R}^{\dagger}=
i \mbox{tr} \left\{\bar{\psi}(1\pm\gamma_{5})\psi
\bar{\psi}(1\pm\gamma_{5})\psi\right\}
\label{ad41}
\end{equation}
and therefore non-vanishing vacuum expectation values of these quartic
fermion operators may develop  irrespective of the realization of
the discrete axial symmetry $\tilde{R}$. In turn, such condensates
cannot be used as order parameters to characterize the
different phases of the discrete axial symmetry.

In addition to this discrete axial symmetry, the full system exhibits the
discrete residual gauge symmetry $S$ which has been explicitly constructed
for SU(2) (eq. (\ref{50})) and SU(3) (eq. (\ref{70b})). Using the results
(\ref{ad13})
and (\ref{ad3}) as well as the invariance of the neutral fermions under
displacements
(eq. (\ref{19h})) and central conjugations  (eq. (\ref{19i})) and  the
transformation  properties
under permutations of the color basis (eq. (\ref{19j})) we obtain
\begin{eqnarray}
\mbox{SU(2):}\quad S\tilde{R}S^{\dagger} &=&-\tilde{R}\nonumber \\
\mbox{SU(3):}\quad S\tilde{R}S^{\dagger} &=&\,\ \tilde{R}\ .
\label{ad18}
\end{eqnarray}
In SU(2), the discrete axial symmetry $\tilde{R}$ is ``anomalous'', i.e.,
this symmetry cannot be realized simultaneously with the discrete residual
gauge symmetry (this is a new example of the {\em global} anomaly!).
``Gauge invariant'', stationary states are therefore
two-fold degenerate with  $\tilde{R}$ connecting these states. Furthermore
the condensate operators develop in general non-vanishing expectation values.
By contrast, in SU(3), the discrete residual gauge symmetry and
axial symmetry can be
realized simultaneously, i.e., stationary states can be labeled by two
quantum numbers characteristic for these two symmetries. In general, the
system does not exhibit degeneracies, and condensates (of operators quadratic
in the fermion fields) are not present.
As in the case of anomalous continuous symmetries, the anomaly of the
discrete $\tilde{R}$ transformation can be cured by supplementing the
fermionic operators by appropriate gauge field operators. This is achieved
easily by replacing the axial charges $(\tilde{Q}_{5})_{ik}$ in the definition
of $\tilde{R}$ by $(Q_{5})_{ik}$
(cf. eqs. (\ref{82c}), (\ref{82e}))
\begin{equation}
R=\exp i\pi\left[\sum _{i,k(i<k)}\left(
(Q_{5})_{ik}+Q_{ik}\right)+\sum_{p=1}^{N-1}
\sum_{k\ge 0}c_{k}^{p\,\dagger}c_{k}^{p}\right]
=\delta\tilde{R} \ .
\label{ad24}
\end{equation}
By construction $R$ is invariant under displacements (\ref{19h})
and central conjugations (\ref{19i})
\begin{equation}
T_{D}(\vec{k})R T_{D}^{\dagger}(\vec{k})=R \quad,\quad
T_{C}(n)R T_{C}^{\dagger}(n)=R \ .
\label{ad31}
\end{equation}
 As in the case of anomalous continuous symmetries,
inclusion of gauge degrees of freedom via $\delta$
in the definition
of $R$ results in a non-vanishing commutator
\begin{equation}
RHR^{\dagger} \neq H
\label{ad32}
\end{equation}
and therefore $R$ is not a symmetry transformation of SU(N)
QCD$_2^{\rm adj}$. This operator is nevertheless useful in clarifying
the general symmetry properties for SU(N). An elementary calculation
shows that the operator $\delta$ acting on the gauge degrees of freedom
transforms as
\begin{equation}
T_{D}(\vec{k})\delta T_{D}^{\dagger}(\vec{k})=\delta \quad,\quad
T_{C}(n)\delta T_{C}^{\dagger}(n)=(-1)^{n\left(N-1\right)}
\delta \ .
\label{ad33}
\end{equation}
Given the invariance of $R$ (eq. (\ref{ad31})) and the invariance of
$\tilde{R}$ under permutations of the color basis (\ref{19j}),
the appropriate relevant SU(N) symmetry transformation $S$ constructed
with these building blocks transforms $\tilde{R}$
as
\begin{equation}
S\tilde{R}S^{\dagger}= (-1)^{n\left(N-1\right)} \tilde{R} \ .
\label{ad34}
\end{equation}
Thus our result derived in detail for SU(3) is generally valid for
SU(N) with $N$ odd. In these systems residual gauge symmetries and
the discrete axial symmetry are simultaneously realized by the stationary
states. No degeneracy occurs as a result of a conflict between discrete
residual gauge and axial symmetries and in general no condensates are
formed. For even $N$ in general the axial symmetry is anomalous
unless the construction of the ``relevant'' symmetry $S$
would not involve the ``elementary'' choice $n=1$ (or more generally
$n$  odd) of the parameter specifying the central conjugations.
(We cannot rule out this possibility at present, since we have not
yet constructed the symmetry operator $S$ for $N>3$.) Assuming
for the moment that $S$ does involve  an odd value of $n$,
we conclude that the stationary states are degenerate. The
transformation property
\begin{equation}
S\tilde{R}S^{\dagger}= -\tilde{R}
\label{ad36}
\end{equation}
implies that the stationary state $|E,z\rangle$ with
\begin{equation}
S|E,z\rangle = z|E,z\rangle
\label{ad37}
\end{equation}
is degenerate with $|E,-z\rangle$. Thus in SU(N) with $N$ even, all the
stationary states might be two-fold degenerate and correspondingly
develop condensates. This result also shows that the relation (\ref{ad36})
cannot be true for odd $N$, where $z$ and -$z$ cannot both belong to the
spectrum of $S$.

Obviously these general results encompass those of the weak coupling
limit. In particular, we confirm the degeneracy of the SU(2)
ground-state with the concomitant formation of a condensate beyond
weak coupling, while the threefold degeneracy of the ground state
in SU(3) (cf. eqs. (\ref{80}), (\ref{81})) is revealed as being due to
symmetries which
do not persist beyond weak coupling. The residual gauge symmetry $S$ is
still present and realized and allows one to  characterize stationary states by
the corresponding  eigenvalue $z$. However as a result of the
 Coulomb interaction these states  must be expected to split energetically.
Concerning the issue of a possibility of an SU(3) condensate, our
discussion only rules out an anomalous symmetry as the origin for a
condensate associated with the fermionic bilinear operators (\ref{ad5}).
Such condensates nevertheless may appear ``dynamically'', like the
chiral condensate in QCD$_{4}$. As emphasized above, appearence
of the quartic condensate (cf. eq. (\ref{ad41})) in the weak-coupling limit
of SU(3) is not
associated with a particular realization of the discrete axial symmetry.
For reasons of continuity, we can  conclude that
this ``quartic'' condensate persists beyond the weak coupling
limit.

We finally comment on  extensions of the local symmetry considerations
beyond the weak coupling. Strictly speaking there is no continuous axial
symmetry which would be broken by effects of regularization of associated
currents. Unlike the axial anomaly of QCD or of the Schwinger model,
the {\em local} axial anomaly of QCD$_2^{\rm adj}$ is a valid concept
only in the weak
coupling limit. It is of interest to repeat the calculation leading to eq.
(\ref{anomaly}) when including the Coulomb interaction. The structure of the
additional term
which arises when commuting the axial charge with $H_{C}$  suggests to
introduce for SU(2) the following ``time-component'' of the vector
potential
\begin{equation}
\alpha_{0}^{ij}(x)=-\frac{g}{L}\sum_{n=-\infty}^{\infty}\int_{0}^{L}
dy\left(1-\delta_{ij}\delta_{n 0}\right)
\frac{\rho^{ij}\left(y\right)}{\left(\frac{2\pi
n}{L}+
\left(i-j\right)a^3\right)^{2}}e^{2i\pi n\left(x-y\right)/L} .
\label{ad10}
\end{equation}
With the help of  $\alpha_{0}$ the time derivative in the continuity equation
is replaced by a covariant derivative
\begin{equation}
\partial ^{0}J_{05}^{3} \left(x\right) -g \epsilon^{ab3}\frac{1}{2}
\left\{J_{05}^{a} \left(x\right), \alpha _{0} ^{b}\left(x\right) \right\}
+\partial ^{1}J_{15}^{3} \left(x\right)=
\frac{g}{2\pi L}\left(e^{3}+e^{3\,\dagger}\right) .
\label{ad11}
\end{equation}
Here we have extended the definition (\ref{82g}) to currents which
have off-diagonal color components.
In accordance  with our above discussion, this result displays the
two different sources for the non-conservation of the axial current.
In addition to the non-vanishing divergence of the axial
current arising from the anomaly (the electric field term
$e^{3}$), the  current is manifestly not conserved
by the appearence of a ``covariant'' time derivative ($\propto g$).

\section{Summary}

Our investigation of QCD$_2^{\rm adj}$ has focused on symmetry properties
of this model field theory and their implications for the vacuum structure
of these models. General topological arguments exhibit the center symmetries,
i.e., $Z_{N}$ symmetries in $SU(N)$ to be the only relevant gauge related
symmetries. These residual gauge symmetries are present only if the
Yang-Mills field is coupled to fermions in the adjoint representation.
 The difference in homotopy properties of $SU(N)/Z_{N}$, the manifold relevant
if fermions are in the adjoint representation, and $SU(N)$ in the case
of fundamental fermions
reveals the presence or absence of these residual symmetries. The nature
of the realization of the symmetries and consequences for the vacuum structure
such as the possible formation of condensates remain unspecified by such
general
topological reasoning. In gauge theories, the connection between
general symmetry
properties and their dynamical implementation is
a remote one.
Topological properties are most easily discussed in a formalism involving
redundant gauge fields; the dynamics on the other hand may be more conveniently
described in terms of unconstrained, physical degrees of freedom. It has been
the purpose of this work to investigate this rather intricate connection
between general topological properties and specific dynamical realizations
in the context of QCD$_2^{\rm adj}$.

We have chosen to perform the detailed dynamical study in the canonical
framework of the Weyl gauge.
We have imposed furthermore periodic boundary conditions, i.e., gauge and
matter fields live on a spatial circle. In this way, the
singular infrared properties are kept under control.
In  the canonical formalism, the redundant gauge
fields are eliminated by explicitly implementing the Gauss law constraint.
The unconstrained, physical degrees of freedom are the fermion fields and,
for SU(N), $N-1$ gauge degrees of freedom  which can be interpreted as zero
momentum neutral gluons. Most important with regard to the topological
properties is the peculiar form of the electric field energy of these
left-over gluons. In the course of eliminating the gauge fields the
kinetic energy of these gluons acquires a non-trivial Jacobian in much the
same way as the kinetic energy of a quantum mechanical particle
moving on the surface of a sphere does. By appropriate definition
of a ``radial''
wave function the kinetic energy can be transformed to the standard cartesian
form supplemented however by the constraint on the   wave function to vanish
wherever the Jacobian vanishes. In this way, the configuration space
of the gauge degrees of freedom becomes a manifold with boundaries. For
SU(2), this manifold is a compact interval, and an equilateral triangle
for SU(3). Along
the boundaries, radial wave function and Jacobian vanish.
Much of the characteristics of the dynamics of QCD$_2$ is due
to these topological
properties. For instance, the different dynamics of
SU(2) QCD$_2$ and the Schwinger-model, with their
common one-dimensional configuration space of the residual gauge
degree of freedom, can be traced back to a large extent to the topological
difference between a compact interval and a circle. (In electrodynamics,
the electric field energy appears without further redefinition of the
wave function in cartesian form.) The physics of a quantum mechanical
particle moving on a circle or in a periodic potential is significantly
different from that
of a particle enclosed e.g. in an infinite square well.
This difference is indicative
of what happens in two-dimensional electrodynamics and SU(2) chromodynamics,
respectively.

The definition of the configuration space of the gauge degrees of
freedom of QCD$_2$ is independent of the characteristics of the
matter fields. The color structure of the matter fields is however
relevant as far as the existence of symmetries acting on  the combined
configuration space of fermion fields and gauge degrees of freedom
 is concerned. We have analyzed in detail these symmetries. As one of
the main results of our studies we have explicitly constructed the residual
gauge symmetries for QCD$_{2}^{\rm adj}$. For SU(2), these symmetry
transformations are reflections of the gauge variable at the midpoint
of the interval defined by the zeroes of the Jacobian accompanied
by a charge conjugation of the fermion fields. In SU(3),
the symmetry transformations consist of a rotation of the fundamental
equilateral triangle by $2\pi/3$ with concomitant color rotations
of the fermions.  The mere presence of these residual gauge symmetries
would have  very  few consequences as far as the spectrum or the structure
of the vacuum  is concerned, were it not for the presence of yet
another symmetry
in this class of theories.  QCD$_2^{\rm adj}$ exhibits in addition to the
gauge symmetries a discrete axial symmetry, i.e., the Hamiltonian is invariant
under a separate change of sign of the totality of right- or of left-``handed"
fermions. The interplay of these two discrete symmetries  is reflected in
spectrum and ground state properties. In particular, these systems provide
an example of  an anomaly in discrete symmetries.  For SU(2N), there is the
possibility that the
reflection symmetry is anomalous, in which case the stationary states cannot
be simultaneously
``gauge invariant'' and of definite chirality. As a consequence, the ground
state is expected in general to be
degenerate and to develop a condensate associated with the standard scalar or
pseudoscalar density. For SU(2N+1) no conflict between the two
symmetries arises
and in general neither degeneracies  occur nor do  scalar or pseudoscalar
densities develop expectation values.

A more specific characterization of the dynamics is possible
in the small  interval
or equivalently the weak coupling limit. Most of our detailed
investigations which
finally have led to the above discussed  general results have been
performed in
this limit. For weak coupling an adiabatic treatment with the gauge
degrees of freedom
representing the slow and the fermions representing the fast degrees
of freedom
is possible. This adiabatic approximation has allowed us to study
very explicitly
the symmetry properties  incorporated in the adiabatic potentials.
The reflection symmetry  in the SU(2) case or  the 3-fold
discrete rotational symmetry of SU(3) QCD$_2^{\rm adj}$ is manifestly
exhibited by the
corresponding adiabatic potential.  This explicit construction
reflects in particular
in a very intuitive way the difference in symmetry arising if the gauge
degrees of freedom are
coupled  to fermions in either  the fundamental or the
adjoint representation.

In the adiabatic picture the formation of condensates
is connected with the overlap
of wave functions of the gauge degrees of freedom associated
with different  fermionic
vacua.
The study of  this overlap has revealed  another interesting
consequence of the
presence of the Jacobian  in the electric field energy. The vanishing
of  Jacobian and
``radial'' wave function occur when the values of gauge degrees
describe pure
gauges and therefore coincide with the minima of the potential energy. As a
consequence, the
modified kinetic energy  forces the system off  the classical
equilibrium position and
thereby enhances the ``tunneling'' probability to other configurations.
 In general, the calculations in the adiabatic, weak coupling limit
are in full agreement with the exact
results and
illustrate the rather surprising odd-even effect  of  condensate formation
in SU(N).
Finally, we have shown that in the weak coupling limit, the axial
symmetry is extended to a continuous symmetry. The presence of continuous
symmetries opens the possibility to study properties
associated with the anomaly using the classical tools. In particular,
we have formulated index theorems associated with the spectral
flow of fermion levels
and could thereby provide further qualitative insights into the symmetry
properties of QCD$_2^{\rm adj}$.

The analysis of the quantum mechanics of the vacuum state of the model
at hand carried
out above gives us a new understanding of the ``condensate" problem.
However, the discrepancy
between fermionic approaches like the present one and bosonization
techniques remains to be clarified.
Our results yield no hint on the bilinear condensate in
the SU(3) case while the bosonization arguments seemingly
indicate that it develops.
If this is indeed true, such a condensate must have a dynamical origin
not directly related
to the gauge symmetry or the discrete chiral symmetry of the theory
and should not
be present in the limit $gL \ll 1$; only then, there would be no obvious
contradiction with our investigation.

We conclude by emphasizing those issues which beyond the particular
structure of QCD$_2^{\rm adj}$ might be of relevance for gauge theories
in higher dimensions. The appearence of Jacobians and centrifugal barriers
which has been of such crucial importance for the structure of the gauge
fixed theory is clearly not limited to gauge theories
in one spatial dimension.
Physical and
unphysical degrees of freedom in non-abelian theories
cannot be expected to simply factorize as in QED,
and this complication does not depend on
the number of space dimensions.
Indeed the corresponding modifications of the electric
field energy have been
found  in a variety of gauges \cite{Christ,GOJA,BAFH,HAJO}.
In most cases, an explicit evaluation of the corresponding
Jacobian is missing. However, irrespective of the detailed
structure of the Jacobians,
their zeroes effectively introduce boundaries into the
infinite dimensional configuration
space of the corresponding theory. Therefore the issue of
realization of symmetries in
the presence of constraints on the wave functional must come
up also in gauge theories
in higher dimensions.  Furthermore,  in the axial gauge
representation of  QCD$_{4}$,
in which the Jacobian can be evaluated explicitly,
the pure gauges, i.e., gauge field
configurations corresponding to vanishing magnetic fields are seen  to be
located on the ``hyperplanes'' of  vanishing Jacobians. Thus an appropriately
defined radial
wave functional is forced to vanish at the classical equilibrium points and
the mechanism of enhanced tunneling processes  which we found  in the one
dimensional case will be effective in higher dimensions too. Finally our
discussion of
anomalies and associated index  theorems should be relevant for gauge
theories in higher dimensions. In the context of QCD$_2^{\rm adj}$ our
discussion
was necessarily restricted to the weak coupling limit. In QCD$_{4}$
on the other hand,
the corresponding continuous axial symmetries and therefore the
phenomenon of ``non-abelian'' anomalies (cf. \cite{Morozov})
persist beyond  the small
volume limit.
In particular it appears promising to extend to higher dimensions our method
of expressing, within a gauge fixed
formulation,  these ``non-abelian''anomalies as abelian anomalies of
the color neutral axial
currents, i.e axial currents  associated with the SU(N) Cartan subalgebra.

\vspace{1.0cm}

\noindent
{\bf Acknowledgements}
\vskip 0.3cm

\noindent
M. S. would like to thank his colleagues from the Institute of Theoretical
Physics,
University of Erlangen-N\"urnberg, where this work was done, for kind
hospitality. Useful discussions with Y. Hosotani, A. Smilga, A. Vainshtein, M.
Voloshin, and K. Yazaki
are gratefully acknowledged.
This work was supported in part by the Alexander von Humboldt
Stiftung
and by DOE under the grant number
DE-FG02-94ER40823 and under the cooperative research agreement
DE-FC02-94ER40818. F. L. and M. T. are supported by the Bundesministerium
f\"{u}r Forschung und Technologie.

\newpage
\noindent
{\large {\bf Figure captions}}

\vspace{0.5cm}

\noindent Fig. 1.
Topology in the space of gauge fields in QCD.
1$a$. A circle in the space of the gauge fields in the $K$ direction.
The length of the circle is 1. The vertical lines indicate the strength of
a potential acting on the effective degree of freedom living on the
circle. If we unwind the circle onto a line we get the picture
of Fig. 1$b$.

\vspace{0.5cm}

\noindent Fig. 2.
An effective potential energy for the degree of freedom living on a
circle in the twisted two-flavor Schwinger model. The point $O'$ is
{\em not} the gauge image of $O$. The true vacuum state is the
symmetric or antisymmetric linear combination of the states
concentrated near $O$ and $O'$.

\vspace{0.5cm}

\noindent Fig. 3.
Topology in the space of fields in QCD$_{2}^{\rm adj}$ with the gauge
group SU(2).

\vspace{0.5cm}

\noindent Fig. 4.
Elementary cell and fundamental domain of the gauge field
variables $a^p$ as defined in the text. a) SU(2), b) SU(3)
gauge group. The thick lines delimit the fundamental
domains. The dashed lines in b) and the midpoint of the
interval $[0, \pi]$ in a) separate regions within the fundamental domain
which are
gauge copies under large gauge transformations.

\vspace{0.5cm}

\noindent Fig. 5.
Adiabatic potential for the gauge degree of freedom in the SU(2) case,
cf. eq. (\ref{41}). Also shown is the ground state wavefunction
$\tilde{\Phi}_{I}(a)$, eq. (\ref{43}), in one of the classical minima,
for the value $gL=1/3$ (arbitrary units).

\vspace{0.5cm}

\noindent Fig. 6.
Fundamental domain of the gauge field variables in SU(3).
The 4 subtriangles $I - IV$ are the regions in
which the Dirac sea has a particular filling. The numbers
in parentheses stand for $(M_{12}, M_{13}, M_{23})$, the
corresponding Fermi levels.

\vspace{0.5cm}

\noindent Fig. 7.
Contour plot of the adiabatic potential $U(a^3,a^8)$ for SU(3),
eqs. (\ref{67}) -- (\ref{70}).
The axes are the same as in Fig. 6. This figure is complemented by
Fig. 8 for the sake of clarity.

\vspace{0.5cm}

\noindent Fig. 8.
Adiabatic potential $U(a^3,a^8)$ for SU(3), eqs. (\ref{67}) -- (\ref{70}).
Shown are cuts through the potential surface along the dashed lines
denoted by a, b, c in Fig. 7.

\end{document}